\documentclass[journal]{IEEEtran}
 
\usepackage{graphicx}
\usepackage{subfig}

\usepackage{amsmath}
\usepackage{graphicx}
\usepackage{multicol}
\usepackage{amssymb}
\usepackage{subfig}
\usepackage{lineno}
\usepackage{makecell}
\usepackage{indentfirst}
\usepackage{bm}
\usepackage{color}
\usepackage{cite}
\usepackage{epstopdf}
\usepackage[noend]{algpseudocode}
\usepackage{algorithmicx,algorithm}
\usepackage{amssymb}
\usepackage{xcolor}
\usepackage{enumerate}
\usepackage{changes}
\usepackage{mathtools}

\newtheorem{mypro}{{Proposition}}
\newtheorem{mylem}{{Lemma}}
\newtheorem{myassu}{{Assumption}}
\newtheorem{mycor}{{Corollary}}

\begin{document}
\title{Cross-Validated Tuning of Shrinkage Factors for MVDR Beamforming Based on Regularized Covariance Matrix Estimation}

\author{Lei~Xie,~
        Zishu~He,~
	   Jun~Tong,~
	   Jun Li and Jiangtao~Xi
\thanks{L. Xie, Z. He and J. Li  are with the School of Information and Communication Engineering,
University of Electronic Science and Technology of China, Chengdu 611731,
China (e-mail: leixie123@hotmail.com).}
\thanks{J. Tong and J. Xi are with the School of Electrical, Computer, and Telecommunications
Engineering, University of Wollongong, Wollongong, NSW 2522, Australia.}
\thanks{This work was supported by the National Natural Science Foundation of China under Grants 61671139 and
61771095.}}

\maketitle
\begin{abstract}
This paper considers the regularized estimation of covariance matrices (CM) of high-dimensional (compound) Gaussian data for minimum variance distortionless response (MVDR) beamforming. 
Linear shrinkage is applied to improve the  accuracy and condition number of the CM estimate for low-sample-support cases. We focus on data-driven techniques that automatically choose the linear shrinkage factors for shrinkage sample covariance matrix ($\mathrm{S}^2$CM) and shrinkage Tyler's estimator (STE) by exploiting cross validation (CV).
We propose leave-one-out cross-validation (LOOCV) choices for the shrinkage factors to optimize the beamforming performance, referred to as $\mathrm{S}^2$CM-CV and STE-CV. The (weighted) out-of-sample output power of the beamfomer is chosen as a proxy of the beamformer performance and concise expressions of the LOOCV cost function are derived to allow fast optimization. For the large system regime, asymptotic approximations of the LOOCV cost functions are derived, yielding the $\mathrm{S}^2$CM-AE and STE-AE. 
In general, the proposed algorithms are able to achieve near-oracle performance in choosing the linear shrinkage factors for MVDR beamforming.
Simulation results are provided for validating the proposed methods.
\end{abstract}


\IEEEpeerreviewmaketitle

\section{Introduction}
\IEEEPARstart{T}{he minimum} variance distortionless response (MVDR) beamformor is a well-established technique in array signal processing \cite{1449208}.  
Example applications include wireless communications, synthetic aperture radar imaging, medical imaging, and space-time adaptive processing (STAP) in radar \cite{godara1997application, habets2009new, byrne2015time, 479429}. 
Assuming that the steering vector of the signal of interest (SOI) is known and exact covariance matrix (CM) of the disturbance consisting of interference and noise  is available,  the MVDR beamformer minimizes the output power of the disturbance while providing a fixed response toward the SOI at the same time. It thus maximizes the signal-to-interference-plus-noise ratio (SINR) at the output of the array. 
However, the exact CM is always unavailable in practice and CM estimation is a key task for the MVDR beamforming. 

CM estimation is also key to many other problems in target detection, direction of arrival (DOA) estimation, sidelobe cancellation, and financial engineering \cite{4101326,135446,bickel2008regularized,6263313, 4104190,Markowitz,9097436,bouchaud2009financial}. 
The most common {CM} estimator is the sample covariance matrix (SCM) \cite{4103115}, which is the maximum likelihood estimator (MLE) of the CM of a Gaussian signal.
However, SCM has two main drawbacks: the inherent inadequacy to the data scarcity and the lack of robustness to outliers\footnote{Following \cite{zeng2013ell}, here an outlier refers to a sample with a high power that has a low probability of occurrence under a nominal Gaussian model of the data.} 
or heavy-tailed distribution of samples.

SCM requires an abundant number of samples to achieve satisfactory performance. 
For beamforming applications, the number of samples need to be twice the dimension to achieve performance within 3 dB from the optimal solution, which is known as Reed, Mallett and Brennan (RMB) criterion \cite{4101326}. 
In order to address the deficiency of samples, algorithms improving the SCM have been proposed, such as those exploiting the structural information of the CM \cite{892662,6166345,6809931}. Another class of algorithms employing linear  shrinkage (regularized) estimators 
has been investigated in \cite{506799,schafer2005shrinkage,8030440, ollila2019optimal,1561576,5417174, 6951419,BUN20171}. 
These algorithms generally  estimate the CM by shrinking an estimate $\mathbf{\Sigma}$ of the CM  toward a well-conditioned target matrix $\mathbf{T}$ with a low condition number\footnote{The condition number of a positive-definite Hermitian matrix is defined as the ratio of the maximum and minimum eigenvalue.}.  There are various choices for $\mathbf{\Sigma}$ and  $\mathbf{T}$. 
Besides the identity or diagonal matrices, the shrinkage target $\mathbf T$ may be set by exploiting  uncertainty models \cite{besson2005performance}, knowledge about the environment \cite{4014427,4490277}, past covariance matrix estimates \cite{1593336}, etc. 
Even multiple shrinkage targets can be applied when distinct guesses of the true CM are available \cite{6935094}.
Introducing shrinkage to SCM yields the shrinkage sample covariance matrix ($\mathrm{S}^2$CM). Similar to the SCM, the $\mathrm{S}^2$CM exhibits poor performance in the presence of outliers.

To address the outliers, one class of approaches is to censor the training samples with the aim to exclude outliers from the CM estimation \cite{6825699,7887259,8401913}. Then the SCM can also provide considerable performance with the censored samples.
 Another class of methods is the robust estimators which have been developed for the wide class of complex elliptically symmetric (CES) distributions \cite{Huber1964Robust, doi:10.1080/01621459.1974.10482962,10.2307/2241079,10.2307/2957994,4359541,6512056,6636083, 6557488, 6630105}.  {These estimators weight the samples by an adaptive factor and aim to reduce the impact of the outliers.} A review on CES with applications to array processing is provided in \cite{6263313}.  In particular, an important subclass of CES distributions, known as compound Gaussian models, has been applied for describing the disturbance (e.g., interference, clutter or noise) returns in radar applications \cite{1055076,6263313,8052571}.  
Similar to the $\mathrm{S}^2$CM for Gaussian case, shrinkage has been applied to the Tyler's estimator \cite{10.2307/2241079} for estimating CM, which results in the shrinkage Tyler's estimator (STE). A major feature of the STE is that it exists even when the number of samples is less than their dimensions, which is in contrast to most conventional robust estimation methods without shrinkage.

When applying the linear shrinkage-based CM estimators to MVDR beamforming, the ultimate  performance depends heavily on the choice of the shrinkage factors.  There are extensive studies    
\cite{LEDOIT2004365,5484583,6913007,ollila2020mestimators,5743027,TONG2018223,xie2021regularized} 
dealing with the choice of shrinkage factors that optimize the estimation of the CM itself, e.g., based on the minimum mean-squared-error (MMSE) criterion. While they may apply to various applications, those techniques do not target the optimization of the performance of MVDR beamforming.  
When the shrinkage target is an identity matrix, the problem is equivalent to choosing the diagonal loading factor (DLF) for diagonally loaded beamformers. 
In \cite{carlson1988covariance}, the DLF is chosen empirically based on the analysis of the noise eigenvalue spread, where the knowledge of the noise level is assumed and utilized. 
In \cite{ma2003efficient}, the DLF is chosen based on the sample standard deviation (SDD) of the diagonal entries of the SCM, assuming that the diagonal entries of the true SCM are equal and the SDD measures the accuracy of the SCM which varies with the sample support.  In \cite{li2003robust}, DL is studied as a solution to robust adaptive beamforming involving mismatched steering vector (SV), and the DLF is calculated using the knowledge of the ellipsoidal uncertainty set describing the inaccurate SV.  
Similar scenarios are studied in \cite{vorobyov2003robust, vincent2004steering}, where an uncertainty model of the SV is adopted, which is in turn used to determine the DLF. 
In general, these solutions require certain user parameters to be given in order to determine the DLF. There are also significant interests in determining the DLF from the training samples only, free of user parameters. Random matrix theory (RMT) has been successfully applied to derive the DLF in \cite{1561576}, which can achieve near-oracle performance in the large-system regime. Furthermore, in \cite{8272485}, MVDR beamformers based on nonlinear shrinkage of the eigenvalues of the SCM is designed by exploiting the spiked model of the covariance matrix and RMT. The solution is applicable to Gaussian disturbances when the number of spikes and noise variance are known.  The aforementioned methods are based on SCM and essentially shrink its eigenvalues to improve the MVDR.  Clearly, their performance depends on the accuracy of the estimation of the principal eigenvectors of the true CM, which are generally sensitive to outliers and heavy-tailed distributions. Shrinkage with non-diagonal targets, which modifies also the eigenvectors of the resulting CM estimate, may improve the performance as compared to DL. Tridiagonal loading is considered in \cite{zhang2019simple}, where the loading level is determined based on the coarse estimate of the desired signal power. More general non-diagonal shrinkage targets are considered in \cite{4490277} where a data-driven general linear combination (GLC) approach is proposed to choose the shrinkage factors. The case with non-Gaussian disturbances is less studied. In \cite{ELKHALIL}, the output performance of the MVDR beamformer with RTE is analyzed while \cite{8082486} proposes approaches for choosing the shrinkage factors of RTE based on RMT.

This paper investigates the tuning of the shrinkage factors for linear shrinkage covariance estimators for optimizing the disturbance suppression performance of  MVDR beamforming for both Gaussian and compound Gaussian disturbances. We target data-driven techniques to automatically tune the linear shrinkage factors for $\mathrm{S}^2$CM and STE, which should be free of user parameters. This is achieved by exploiting leave-one-out cross-validation (LOOCV). 
Although cross validation (CV) has been used extensively in tuning parameters in signal processing,  machine learning and statistics \cite{arlot2010survey}, it is critical to choose an appropriate cost function to not only optimize performance but also simplify its computational cost for fast or even closed-form solution to the parameter. In this paper, we address this challenge for MVDR beamforming and investigate the performance of the resulting shrinkage factor choice. 
 Our contributions are as follows:
\begin{itemize} 
\item 

For $\mathrm{S}^2$CM with a general shrinkage target matrix, we use the average out-of-sample output power of the beamformer as an effective CV cost. In order to remove the obstacle of the high computational complexity for general CV procedures, a concise expression is derived to allow for its fast evaluation, based on which the best shrinkage factors are chosen. We refer to the resulting CM estimator as $\mathrm{S}^2$CM-CV.

\item Exploiting the fact that Tyler's estimator can be interpreted as a SCM of weighted samples, we then extend  $\mathrm{S}^2$CM-CV to compound Gaussian disturbances,   
yielding the STE-CV methods. Since the STE need to be found iteratively, the complexity will be prohibitively high if standard CV is applied. To address this concern, the compound Gaussian samples are weighted by using plug-in estimates of the true CM before the CV procedure. Depending on whether a fixed or adaptive plug-in CM estimate is used, two methods, i.e.,  STE-CV-I and STE-CV-II, are obtained to yield different performance and complexity. These methods provide data-driven alternatives to tune the STE for MVDR beamforming.

\item We derive asymptotic estimators $\mathrm{S}^2$CM-AE and STE-AE  that well approximate the $\mathrm{S}^2$CM-CV and STE-CV methods in the high-dimensional regime, for Gaussian and non-Gaussian disturbance, respectively. We show that $\mathrm{S}^2$CM-AE is equivalent to the solution of \cite{1561576} for the Gaussian case with identity targets, which was originally derived using RMT. The proposed asymptotic estimators extend the solution of \cite{1561576} to non-identity targets and non-Gaussian disturbances.

\item We analyze the computational complexities of the proposed methods and demonstrate their effectiveness by numerical studies with different shrinkage targets, sample sizes,  and disturbance distributions. The proposed methods are shown to be able to approach the oracle performance in general. Compared with several existing methods, the proposed estimators can exhibit wider applicability or better performance for MVDR beamforming.

\end{itemize}

The remainder of this paper is organized as follows. Section II establishes the mathematical model. Section III and IV introduce the proposed $\mathrm{S}^2$CM-CV (and its extension $\mathrm{S}^2$CM-AE) and STE-CV (and its extension STE-AE) schemes for Gaussian and compound Gaussian cases, respectively. 
Section V presents simulation examples to show the performance. Finally, Section VI gives the conclusions.

\section{Problem Formulation}

\subsection{Mathematical Model}
Consider a detection problem with the length-$N$ observed signal 
\begin{equation}
\label{sigmodel}
 	 \mathbf{y}  = a\mathbf{s} + \mathbf{c} ,    
\end{equation} 
where $\mathbf s$ is the known template of the signal-of-interest (SOI) with $\Vert \mathbf s \Vert^2=N$, $a$ is an unknown complex gain, and $\mathbf{c} $ stands for the disturbance ({may consist of interference, clutter or noise}). 
In robust statistics,  $\mathbf c$ can be described using heavy-tailed distributions and outlier contamination \cite{Huber1964Robust}.  
In many radar scenarios, experimental trials have shown a good fitting of the zero-mean {compound Gaussian model}, an instance of   the complex elliptically symmetric (CES) distributions, to the disturbance measurements \cite{89019,4647137,766928,766939}. 
The received disturbance can be modeled as
\begin{equation}
    \begin{split}
	\mathbf{c}  \triangleq   \sqrt{\tau} \mathbf{x}  =\sqrt{\tau} \mathbf{A} \mathbf{e}  \in \mathbb{C}^{N\times 1},
    \end{split}
\end{equation} 
where the texture $\tau$ is a real-valued and positive random variable (R.V.),  which is independent of the speckle $\mathbf{x}\triangleq \mathbf{A} \mathbf{e}$. 
The circular-symmetric complex Gaussian random vector $\mathbf{e} \in \mathbb{C}^{M\times 1}$ has independent entries with zero mean and unit variance, the deterministic matrix $\mathbf{A} \in \mathbb{C}^{N\times M}$ with $M\geq N$, and $\mathbf{R}=\mathbf{A}\mathbf{A}^\mathrm{H}$ denotes the covariance matrix of $\mathbf{x}$. 
The covariance matrix of $\mathbf{c}$ is thus defined as $\mathbf{R}_c=\mathbb{E}(\tau) \mathbf{R}$.
In order to define $\mathbf{R}$ uniquely, we assume $\mathrm{Tr}(\mathbf{R}) =N$ and the resultant scaling can be absorbed by $\tau$. The speckle $\mathbf{x}=\mathbf{A} \mathbf{e}$ can also be used to model the zero-mean Gaussian vector.
 {Note that the Gaussian case is a special example of the compound Gaussian one.}
 If $\tau \equiv 1$, i.e., $\mathbf{c}=\mathbf{x}$, the compound Gaussian model  degrades to the ordinary Gaussian one.

The MVDR beamformer provides an effective approach to suppressing the disturbance  $\mathbf{c}$, whose weight  is given by \cite{1449208} 
\begin{equation}
    \begin{split}
	\min_{\mathbf{w}} &\quad \mathbf{w}^{\mathrm{H}} \mathbf{R}_c \mathbf{w}\\
	s.t. &\quad \mathbf{w}^{\mathrm{H}}\mathbf{s}=1
    \end{split}
    \label{MVDRoriginal}
\end{equation} 
where $\mathbf{s}$ is also referred to as the steering vector of the signal of interest.
The optimum weight is computed as 
\begin{equation}
    \begin{split}
	\mathbf{w}=\frac{\mathbf{R}_c^{-1} \mathbf{s}}{ \mathbf{s}^\mathrm{H}    \mathbf{R}_c^{-1} \mathbf{s}}=\frac{\mathbf{R}^{-1} \mathbf{s}}{ \mathbf{s}^\mathrm{H}    \mathbf{R}^{-1} \mathbf{s}}.
    \end{split}
    \label{wopt}
\end{equation} 
This beamformer can minimize the output power of the disturbance under a fixed response to $\mathbf{s}$. 
With (\ref{wopt}),  the mainlobe of the beamformer is set to the direction of the target signal,  {whose output is expected to be distortionless}. Meanwhile, its sidelobes are adapted to the disturbance properties to suppress their contributions to the outputs.

Since $\mathbf{R}$ is always unknown in practice, such a beamformer is unavailable in practice. Its performance  depends not on $\mathbb{E}(\tau)$, but on the exact CM of $\mathbf{x}$, i.e., $\mathbf{R}$. Therefore, in this paper, we only consider the estimation of $\mathbf{R}$. We can substitute an estimate $\mathbf{\Sigma}$ of $\mathbf{R}$  into  (\ref{wopt}) and obtain 
\begin{equation}
    \begin{split}
	\widehat{\mathbf{w}} =\frac{\mathbf{\Sigma}^{-1} \mathbf{s}}{ \mathbf{s}^\mathrm{H}    \mathbf{\Sigma}^{-1} \mathbf{s}}.
    \end{split}
    \label{what}
\end{equation}

The performance of the beamformer can be measured by the output power of the disturbance  \cite{8272485}, which is  a function of $\mathbf{\Sigma}$ and has the form as
\begin{equation}
    \begin{split}
 \mathcal{J}\left( \mathbf{\Sigma} \right)
  \triangleq  \widehat{\mathbf{w}}^\mathrm{H}  \mathbf{R}_c  \widehat{\mathbf{w}}
   =\mathbb{E}(\tau) \frac{\mathbf{s}^\mathrm{H}    \mathbf{\Sigma}^{-1}    \mathbf{R}  \mathbf{\Sigma}^{-1} \mathbf{s}}{ \left(\mathbf{s}^\mathrm{H}   \mathbf{\Sigma}^{-1} \mathbf{s}\right)^2}.
    \end{split}
    \label{cost00}
\end{equation} 
With $\mathbb{E}(\tau) $ being a fixed constant, the cost function (\ref{cost00})  is minimized when $ \mathbf{\Sigma}=\mathbf{R}$ \cite{8272485}.  
However, (\ref{cost00}) is unavailable in practice due to the 
unknown $\mathbf{R}$. In this paper, we target estimating 
(\ref{cost00}) from the training data  so that we can compare and choose MVDR implementations.

\subsection{Covariance (Scatter) Matrix Estimator}
\label{SecMLE}
Let $\mathcal{Y}=\{\mathbf{y}_l\}_{l=1}^L$ be a set of $L$ independent and identically distributed (i.i.d.) samples of $\mathbf c$,
which is modeled as
\begin{equation}
    \begin{split}
    \mathbf{y}_l=\sqrt{\tau_l} \mathbf{x}_l=\sqrt{\tau_l} \mathbf{A} \mathbf{e}_l, l=1,2,\ldots,L.
    \end{split}
\end{equation} 
Note that all $\mathbf{x}_l$ share the same $\mathbf{A}$.
We assume that each sample has a finite power, i.e., $\Vert  \mathbf{y}_l \Vert<\infty$, and all samples are evenly spread out in the whole space \cite{10.2307/2241079,6879466}. 
We estimate the covariance matrix of $\mathbf{x}$, i.e., $\mathbf{R}$, using the secondary data $\mathbf{y}_l, l=1,2,\ldots,L$, 
and then cancels the disturbance component in $\mathbf{y}$. 

If $\tau_l =1$, $\mathbf{y}_l =\mathbf{x}_l$ is the Gaussian samples.
The SCM is a common estimator since it is the maximum likelihood estimator (MLE) for the Gaussian data, which is written as
\begin{equation}
    \begin{split}
	\widehat{\mathbf{R}}_{\mathrm{SCM}}=\frac{1}{L} \sum_{l=1}^{L} \mathbf{x}_l \mathbf{x}_l^\mathrm{H}.
    \end{split}
    \label{SCM}
\end{equation} 
By definition, we have
\begin{equation}
    \begin{split}  
 \mathbb{E} \left(  \mathbf{x} \mathbf{x}^\mathrm{H}    \right) = \mathbf{R}.   
    \end{split}
    \label{EXXRG}
\end{equation} 
This indicates that, for Gaussian samples, SCM is an unbiased estimator for $\mathbf{R}$ with $\mathbb{E} \left( \widehat{\mathbf{R}}_{\mathrm{SCM}} \right) =\mathbf{R}$.

For non-Gaussian data, $\{\mathbf x_l\}$ are unavailable and the SCM of $\{\mathbf y_l\}$ may however perform poorly due to its lack of robustness against outliers.
An important estimator is Tyler's estimator (TE) \cite{tyler1987distribution}, i.e., the solution to the following fixed-point equation
\begin{equation}
    \begin{split}
	 \widehat{\mathbf{R}}_{\mathrm{Tyler}}=\frac{1}{L} \sum_{l=1}^{L} \frac{\mathbf{y}_l \mathbf{y}_l^\mathrm{H}}{\frac{1}{N}  \mathbf{y}_l^\mathrm{H}  \widehat{\mathbf{R}}_{\mathrm{Tyler}}^{-1} \mathbf{y}_l}.
    \end{split}
    \label{Tyler}
\end{equation} 
Tyler's estimator eliminates the influence of $\tau_l$ in the samples and brings robustness against outliers with large $\tau_l$.  
The $M$-functional matrix $\mathbf{R}_{u,\mathrm{Tyler}}$ for Tyler's estimator is defined as a solution of \cite{6263313,ollila2020mestimators}
\begin{equation}
    \begin{split} 
\mathbb{E} \left(  \frac{ N}{\mathbf{y}^\mathrm{H} \mathbf{R}_{u,\mathrm{Tyler}}^{-1} \mathbf{y} }  \mathbf{y} \mathbf{y}^\mathrm{H} \right) =\mathbf{R}_{u,\mathrm{Tyler}},
\end{split}
\label{ERu}
\end{equation} 
where $\mathbf{R}_{u,\mathrm{Tyler}}=\sigma_{u,\mathrm{Tyler}} \mathbf{R}$ \cite{6263313} and $\sigma_{u,\mathrm{Tyler}}$ is a constant. Dividing both sides of this equation by $\sigma_{u,\mathrm{Tyler}}$, we obtain 
\begin{equation}
    \begin{split}   
    \mathbb{E} \left(  \frac{\mathbf{y} \mathbf{y}^\mathrm{H} }{\frac{1}{N}\mathbf{y}^\mathrm{H} \mathbf{R}^{-1} \mathbf{y} }   \right) =\mathbf{R}.   
\end{split}
    \label{ExpTyler}
\end{equation}

\section{ $\mathrm{S}^2$CM Cross-Validated for MVDR Beamforming: Gaussian Distribution}

In this section, we study the shrinkage estimation of CM for MVDR beamforming with Gaussian disturbance. 
The SCM can be ill-conditioned if the number of samples $L$ is comparable to or smaller than the dimension $N$. 
To address this issue, one may apply the following shrinkage SCM, which will be referred to as $\mathrm{S}^2$CM in the following: 
\begin{equation}
    \begin{split}
	\widehat{\mathbf{R}}_{\mathrm{G}} \left(  \rho \right)=(1-\rho)\widehat{\mathbf{R}}_{\mathrm{SCM}} +\rho \mathbf{T},
    \end{split}
    \label{RSCM0}
\end{equation} 
 where $\rho \in [0,1)$ is  the shrinkage factor  and $\mathbf{T}$ denotes the target matrix which is assumed positive-definite and well-conditioned. 
 Clearly, the choice of $\rho$ affects $ \widehat{\mathbf{R}}_{\mathrm{G}} \left(  \rho \right)$. The choice that minimizes (\ref{cost00}) with $\tau=1$ will be referred to as the $\mathrm{S}^2$CM-oracle choice for the MVDR beamformer, i.e.,
\begin{equation}
    \begin{split}
    \rho_{\mathrm{S^2CM-oracle}}=\min_{\rho} \mathcal{J} \left( \widehat{\mathbf{R}}_{\mathrm{G}} \left(  \rho \right)   \right) .
\end{split}
    \label{SSCMoracle}
\end{equation} 
 Since it depends on the unknown covariance matrix $\mathbf R$, $\rho_{\mathrm{S^2CM-oracle}}$ is not accessible in practice. 
This section aims to address the challenge of finding a practical shrinkage factor $\rho$ that approaches $\rho_{\mathrm{S^2CM-oracle}}$ so that the performance of the resultant MVDR beamformer can be optimized.

Recently, the general model selection tool CV has been used to successfully obtain low-cost solutions to shrinkage CM estimation.   
For example, \cite{TONG2018223} derives analytical LOOCV solutions to minimize the MSE of the CM estimation for Gaussian data, 
\cite{8755295} proposes a CV solution to optimize the inverse covariance matrix (a.k.a. precision matrix) estimation, while \cite{7422135, 8030440} derive LOOCV solutions for the application of minimum mean square error (MMSE) filtering. In this paper, we extend these works to  
obtain data-driven and low-complexity approaches for choosing $\rho$ for the MVDR beamformer, by finding a suitable CV cost as an estimate of $\mathcal{J}( \mathbf{\Sigma} )$ in (\ref{cost00}) and its fast evaluation.

\subsection{LOOCV Choice of the Shrinkage Factor}
\label{SecChoiceofG}

 Employing the LOOCV strategy, the $L$ samples $\mathcal{X}=\{ \mathbf{x}_{l}\}_{l=1}^L$ are repeatedly split into two sets. For the $l$th split, the training subset $\mathcal{X}_l$ (with $\mathbf x_l$ omitted from $\mathcal{X}$) is used for producing a covariance matrix estimate 
\begin{equation}
    \begin{split}
\widehat{\mathbf{R}}_{\mathrm{G},l} (\rho)=(1-\rho) \frac{1}{L-1} \sum_{i\neq l} \mathbf{x}_i \mathbf{x}_i^\mathrm{H} +\rho \mathbf{T},
    \end{split}
\end{equation} 
and the remaining  sample $\mathbf{x}_{l}$ is spared for parameter validation.  
 Substituting $\mathbf{\Sigma}$ and $\mathbf{R} $ with $\widehat{\mathbf{R}}_{\mathrm{G},l} (\rho)$ and $\mathbf{x}_l \mathbf{x}_l^\mathrm{H}$ in (\ref{cost00}), we obtain the LOOCV output power for the $l$th split 
 \begin{equation}
    \begin{split}
P_l=\left|  \mathbf{w}_{\mathrm{G},l}^\mathrm{H} \mathbf{x}_l  \right|^2 
    \end{split}
\end{equation} 
as a proxy for (\ref{cost00}), where the MVDR beamformer constructed from $\mathrm{S}^2$CM using $\mathcal{X}_l$ is given by 
\begin{equation}
    \begin{split}
	\mathbf{w}_{\mathrm{G},l}=\frac{  \widehat{\mathbf{R}}_{\mathrm{G},l}^{-1} (\rho) \mathbf{s}}{ \mathbf{s}^\mathrm{H}    \widehat{\mathbf{R}}_{\mathrm{G},l}^{-1} (\rho) \mathbf{s}}=\frac{  \widetilde{\mathbf{R}}_{\mathrm{G},l}^{-1} (\alpha) \mathbf{s}}{ \mathbf{s}^\mathrm{H}   \widetilde{\mathbf{R}}_{\mathrm{G},l}^{-1} (\alpha) \mathbf{s}}.
    \end{split}
    \label{whatGL}
\end{equation} 
The MVDR beamformer $\mathbf{w}_{\mathrm{G},l}$ aims to minimize the response to $\mathbf{x}_l$ which contains no signal component and serves as a validation set independent of the training set $\mathcal{X}_l$. Intuitively, a good choice of $\rho$ leads to a low output power $P_l$.

It can be checked that $\mathbf{w}_{\mathrm{G},l}$  is invariant to scaling on $\widehat{\mathbf{R}}_{\mathrm{G},l}$ and we have
\begin{equation}
    \begin{split}
	\widetilde{\mathbf{R}}_{\mathrm{G},l} (  \alpha )	&=\frac{1}{L} \sum_{i\neq l} \mathbf{x}_i \mathbf{x}_i^\mathrm{H}  +\alpha \mathbf{T}=\widetilde{\mathbf{R}}_{\mathrm{G}} \left(  \alpha \right)-\frac{1}{L}  \mathbf{x}_l \mathbf{x}_l^\mathrm{H} ,
    \end{split}
    \label{tildeRkl1}
\end{equation} 
where $\alpha=\frac{\rho (L-1)}{L(1-\rho)}$ and 
\begin{equation}
    \begin{split}
\widetilde{\mathbf{R}}_{\mathrm{G}} \left(  \alpha \right)=  \widehat{\mathbf{R}}_{\mathrm{SCM}}+\alpha \mathbf{T}. 
\end{split}
    \label{RscmAI}
\end{equation} 
Note that each $\rho$ corresponds to a unique $\alpha$. Though we use $\alpha$ in derivations for notational simplicity, we aim to finally find an optimized $\rho$ to use in (\ref{RSCM0}).

In total, $L$ splits of $\mathcal{X}$ are used and all the training samples are used for validation once. 
Then we construct the following LOOCV function
\begin{equation}
    \begin{split}
 &\mathcal{J}_{\mathrm{CV}}\left( \alpha\right) = \frac{1}{L} \sum_{l=1}^L  \widehat{P}_l=\frac{1}{L} \sum_{l=1}^L   \left| \frac{   \mathbf{s}^\mathrm{H} \widetilde{\mathbf{R}}_{\mathrm{G},l}^{-1}   (  \alpha ) \mathbf{x}_l}{  \mathbf{s}^\mathrm{H}    \widetilde{\mathbf{R}}_{\mathrm{G},l}^{-1}   (  \alpha ) \mathbf{s}} \right|^2.
    \end{split}
    \label{CostGCV00}
\end{equation} 
In order to obtain an appropriate estimate of the optimum  shrinkage factor, we propose to evaluate the LOOCV cost function in (\ref{CostGCV00}) and search for its global minimum, i.e.,
\begin{equation}
    \begin{split}
 {\alpha_{\mathrm{CV}}^{\star} =\arg\min_{\alpha}\mathcal{J}_{\mathrm{CV}} \left( \alpha \right).} 
    \end{split}
    \label{alphaCV000}
\end{equation}

It is easy to see that $L$ matrix inversions\footnote{The computational complexity of inverting an $N \times N$ matrix is $\mathcal{O}(N^3)$.} are involved in (\ref{CostGCV00}) for $\widetilde{\mathbf{R}}_{\mathrm{G},l} (  \alpha )$. The total computational complexity of (\ref{CostGCV00}) is about $\mathcal{O}(K_G  L N^3 )$, where $K_G$ denotes the number of  candidates for  the grid-search of $\alpha$. 
Clearly, a direct implementation of the above LOOCV method is computationally expensive, which shares one of the main bottlenecks of CV. This motivates us to derive concise expressions to reduce the computational cost. 
Applying the matrix inversion lemma, we have
\begin{equation}
    \begin{split}
           \widetilde{\mathbf{R}}_{\mathrm{G},l}^{-1}   (  \alpha )&=\left(  \widetilde{\mathbf{R}}_{\mathrm{G}} \left(  \alpha \right)-\frac{1}{L}  \mathbf{x}_l \mathbf{x}_l^\mathrm{H}  \right)^{-1}\\
           &= \widetilde{\mathbf{R}}_{\mathrm{G}}^{-1}(\alpha)-\frac{\widetilde{\mathbf{R}}_{\mathrm{G}}^{-1} (\alpha) \mathbf{x}_l \mathbf{x}_l^\mathrm{H} \widetilde{\mathbf{R}}_{\mathrm{G}}^{-1}(\alpha)   }{\mathbf{x}_l^\mathrm{H} \widetilde{\mathbf{R}}_{\mathrm{G}}^{-1}(\alpha) \mathbf{x}_l-L }.
    \end{split}
    \label{Rglalphainv}
\end{equation} 
Then we have
\begin{subequations}
\begin{equation}
    \begin{split}
&\mathbf{s}^\mathrm{H} \widetilde{\mathbf{R}}_{\mathrm{G},l}^{-1}   (  \alpha ) \mathbf{x}_l
=\mathbf{s}^\mathrm{H}\widetilde{\mathbf{R}}_{\mathrm{G}}^{-1}(\alpha) \mathbf{x}_l  
\left(  \frac{1}{1-  \frac{1}{L}   \mathbf{x}_l^\mathrm{H} \widetilde{\mathbf{R}}_{\mathrm{G}}^{-1} (\alpha) \mathbf{x}_l }               \right),\\
    \end{split}
    \label{ary0}
\end{equation} 
\begin{equation}
    \begin{split}
&\mathbf{s}^\mathrm{H} \widetilde{\mathbf{R}}_{\mathrm{G},l}^{-1} (\alpha) \mathbf{s}=\mathbf{s}^\mathrm{H}\widetilde{\mathbf{R}}_{\mathrm{G}}^{-1}(\alpha) \mathbf{s} -\frac{\mathbf{s}^\mathrm{H}\widetilde{\mathbf{R}}_{\mathrm{G}}^{-1} (\alpha) \mathbf{x}_l \mathbf{x}_l^\mathrm{H} \widetilde{\mathbf{R}}_{\mathrm{G}}^{-1} (\alpha) \mathbf{s}}{\mathbf{x}_l^\mathrm{H} \widetilde{\mathbf{R}}_{\mathrm{G}}^{-1} (\alpha)\mathbf{x}_l- L }.\\
    \end{split}
    \label{ara0}
\end{equation} 
\end{subequations}
We can thus get the beamformer output at the $l$th split as 
\begin{equation}
    \begin{split}
&\widehat{\mathbf{w}}_l^\mathrm{H} \mathbf{x}_l  = \frac{  \mathbf{s}^\mathrm{H} \widetilde{\mathbf{R}}_{\mathrm{G},l}^{-1}   (  \alpha ) \mathbf{x}_l }{ \mathbf{s}^\mathrm{H}    \widetilde{\mathbf{R}}_{\mathrm{G},l}^{-1}   (  \alpha ) \mathbf{s}} \\
 &= \frac{  \mathbf{s}^\mathrm{H}\widetilde{\mathbf{R}}_{\mathrm{G}}^{-1} (\alpha)\mathbf{x}_l  }{ \mathbf{s}^\mathrm{H}\widetilde{\mathbf{R}}_{\mathrm{G}}^{-1} (\alpha)\mathbf{s}}
 \cdot
\frac{1}{1-\frac{1}{L}  \mathbf{x}_l^\mathrm{H} \widetilde{\mathbf{R}}_{\mathrm{G}}^{-1} (\alpha) \mathbf{x}_l +\frac{1}{L}   \frac{\left|\mathbf{s}^\mathrm{H}\widetilde{\mathbf{R}}_{\mathrm{G}}^{-1} (\alpha) \mathbf{x}_l \right|^2}{\mathbf{s}^\mathrm{H}\widetilde{\mathbf{R}}_{\mathrm{G}}^{-1} (\alpha)\mathbf{s}  }}.
    \end{split}
    \label{wk1lHul0}
\end{equation} 
Substituting (\ref{wk1lHul0}) into (\ref{CostGCV00}), the LOOCV cost function can be rewritten as
\begin{equation}
    \begin{split}
 {\mathcal{J}_{\mathrm{CV}} \left( \alpha \right)= \frac{1}{L} \sum_{l=1}^L  \frac{ \mathbf{s}^\mathrm{H} \widetilde{\mathbf{R}}_{\mathrm{G}}^{-1}\left( \alpha \right) \mathbf{x}_l\mathbf{x}_l^\mathrm{H}   \widetilde{\mathbf{R}}_{\mathrm{G}}^{-1}\left( \alpha \right) \mathbf{s} }{\left(1-\frac{N}{L} \phi_l(\alpha) \right)^2 \left(\mathbf{s}^\mathrm{H}   \widetilde{\mathbf{R}}_{\mathrm{G}}^{-1}\left( \alpha \right)  \mathbf{s} \right)^2 },} 
    \end{split}
    \label{Jcvk1}
\end{equation}
where 
 \begin{equation}
    \begin{split}
    	\phi_l (\alpha)&=\frac{1}{N} \left(\mathbf{x}_l^\mathrm{H} \widetilde{\mathbf{R}}_{\mathrm{G}}^{-1}(\alpha) \mathbf{x}_l -\frac{ \left|\mathbf{s}^\mathrm{H}\widetilde{\mathbf{R}}_{\mathrm{G}}^{-1}(\alpha) \mathbf{x}_l \right|^2
 }{  \mathbf{s}^\mathrm{H}\widetilde{\mathbf{R}}_{\mathrm{G}}^{-1}(\alpha) \mathbf{s}}  \right).
    \end{split}
    \label{phil}
\end{equation} 
Then the proposed algorithm is summarized in Algorithm  \ref{alg_S2CMCV}. 
 Using (\ref{Jcvk1}), the overall computational complexity of   Algorithm \ref{alg_S2CMCV} is $\mathcal{O}(N_G N^3+ N_G L N^2)$ which mainly arises from finding the inverses $\{\widetilde{\mathbf{R}}_{\mathrm{G}}^{-1}(\alpha)\}$  {for different $\alpha$} and multiplying them to vectors $\{\mathbf x_l\}$.

\begin{algorithm}[t] 
\caption{The $\mathrm{S}^2$CM-CV Estimator} 
\begin{enumerate} 
  \item Compute the optimized shrinkage parameter $\rho$ via a numerical search 
  \[\rho_{\mathrm{CV}}^{\star} =\arg\min_{\rho \in (0,1)}\mathcal{J}_{\mathrm{CV}} \left( \alpha(\rho) \right), \]
  where   $\mathcal{J}_{\mathrm{CV}} \left( \alpha \right)$ is defined as (\ref{Jcvk1}) and $\alpha(\rho )=\frac{\rho (L-1)}{L(1-\rho )}$.
  \item Compute the $\mathrm{S}^2$CM-CV  via (\ref{RSCM0}).
\end{enumerate} 
\label{alg_S2CMCV} 
\end{algorithm}

 The computational complexity may be further reduced for the identity target $\mathbf{T} = \mathbf{I}$. 
From (\ref{Jcvk1}), the inverse of $\widetilde{\mathbf{R}}_{\mathrm{G}}\left( \alpha \right)$ is involved for each candidate $\alpha$. 
In order to efficiently evaluate the cost function in (\ref{Jcvk1}) for different $\alpha$, eigendecomposition (EVD) of  $\widehat{\mathbf{R}}_{\mathrm{SCM}}$ (at a cost of $\mathcal{O}(N^3 )$) in (\ref{SCM}) can be computed as 
\begin{equation}
    \begin{split}
		\widehat{\mathbf{R}}_{\mathrm{SCM}}=\mathbf{V} {\mathbf{\Lambda}} {\mathbf{V}}^\mathrm{H},
    \end{split}
    \label{SCMsvd}
\end{equation} 
where
${\mathbf{V}}=\left[ {\mathbf{v}}_{1},{\mathbf{v}}_{2},\cdots,{\mathbf{v}}_{N}    \right]$, $\mathbf{\Lambda}=\mathrm{diag}\left( {\lambda}_{1},{\lambda}_{2},\cdots,{\lambda}_{N}    \right)$ and ${\lambda}_{1}   \geq {\lambda}_{2} \geq \cdots  \geq {\lambda}_{N}\geq 0$. 
Then (\ref{wk1lHul0}) can be rewritten as
\begin{equation}
    \begin{split}
\widehat{\mathbf{w}}_l^\mathrm{H} \hat{\mathbf{x}}_l= 
 \frac{C_{\mathbf{s},\hat{\mathbf{x}}_l}(\alpha) }{ C_{\mathbf{s}}(\alpha)-\frac{1}{L}    C_{\mathbf{s}}(\alpha)C_{\hat{\mathbf{x}}_l}(\alpha)  +\frac{1}{L}  \left|C_{\mathbf{s},\hat{\mathbf{x}}_l}(\alpha)\right|^2 },
    \end{split}
\end{equation}
where
\begin{subequations}
\begin{equation}
    \begin{split}
C_{\mathbf{s},\hat{\mathbf{x}}_l}(\alpha)&=\mathbf{s}^\mathrm{H} \widetilde{\mathbf{R}}_{\mathrm{G}}^{-1}(\alpha) \hat{\mathbf{x}}_l=\sum_{i=1}^N \frac{\left(    {\mathbf{v}}_{i}^\mathrm{H} \mathbf{s} \right)^{*}{\mathbf{v}}_{i}^\mathrm{H} \mathbf{x}_l }{{\lambda}_{i}+\alpha} ,
    \end{split}
\end{equation}
\begin{equation}
    \begin{split}
C_{\mathbf{s}}(\alpha)=\mathbf{s}^\mathrm{H} \widetilde{\mathbf{R}}_{\mathrm{G}}^{-1}(\alpha)\mathbf{s}=\sum_{i=1}^N \frac{\left|    {\mathbf{v}}_{i}^\mathrm{H} \mathbf{s} \right|^{2} }{{\lambda}_{i}+\alpha},
    \end{split}
\end{equation}
\begin{equation}
    \begin{split}
C_{\hat{\mathbf{x}}_l}(\alpha)&=\hat{\mathbf{x}}_l^\mathrm{H} \widetilde{\mathbf{R}}_{\mathrm{G}}^{-1}(\alpha) \hat{\mathbf{x}}_l=\sum_{i=1}^N \frac{\left|    {\mathbf{v}}_{i}^\mathrm{H}  \mathbf{x}_l \right|^{2}  }{{\lambda}_{i}+\alpha} .
    \end{split}
\end{equation}
\end{subequations}
Note that in the above, ${\mathbf{v}}_{i}^\mathrm{H}  \mathbf{x}_l$, $i=1,\cdots, N, l=1,\cdots, L$ are irrelevant to $\alpha$ and thus need to be evaluated only once and reused for different $\alpha$,  {whose computational complexity is about $\mathcal{O}(LN^2)$}.  
Thus (\ref{Jcvk1}) can be computed as
\begin{equation}
    \begin{split}
\mathcal{J}_{\mathrm{CV}} \left( \alpha \right)
=\frac{1}{L} \sum_{l=1}^L   \left| \frac{C_{\mathbf{s},\hat{\mathbf{x}}_l}(\alpha) }{ C_{\mathbf{s}}(\alpha)-\frac{1}{L}    C_{\mathbf{s}}(\alpha)C_{\hat{\mathbf{x}}_l}(\alpha)  +\frac{1}{L}  \left|C_{\mathbf{s},\hat{\mathbf{x}}_l}(\alpha)\right|^2 } \right|^2.
    \end{split}
    \label{JCVefficient}
\end{equation}
 with a computational complexity of $\mathcal{O}(N_G LN )$. 
Then the total computational cost of Algorithm \ref{alg_S2CMCV}  can be reduced to  {$\mathcal{O}(N^3+ L N^2 + N_G LN )$}.

\subsection{ {Asymptotic Estimator}} 
 
This subsection presents an asymptotic estimator (AE), referred to as {$\mathrm{S}^2$CM-AE}, for the large $(N, L)$ regime, which has a more concise expression. 
For identity targets, we will apply results from random matrix theory (RMT)  \cite{6891244,COUILLET201499,8352743} to demonstrate that $\mathrm{S}^2$CM-AE is equivalent to $\mathrm{S}^2$CM-CV 
in the asymptotic regime where both the dimensionality $N$ and number of samples $L$ approach infinity.  

 Let us make the following statistical assumptions on the large-dimensional random matrices under study, unless otherwise stated \cite{6891244,COUILLET201499,8352743}:
\begin{myassu}
\label{assuption1}
\begin{enumerate}[(1)]
\item 
 The entries of $\mathbf{e}_l$, i.e., 
$e_{l,i}, i=1, \cdots, M, l=1,\cdots, L$, are i.i.d. random variables in $\mathbb{C}$ with $\mathbb{E} (e_{l,i})=0$, $\mathbb{E} ( |e_{l,i}|^2)=1$ and $\mathbb{E} ( |e_{l,i}|^{ {8}})< \infty$.
\item 
$c_N\triangleq N/L\to c \in (0,\infty)$ as $N, L\to \infty$. 
\item 
There exist $C_{-}$ and $C_{+}>0$ such that
\[C_{-}<\lim \inf_N  \lambda_{\min}(\mathbf{R})  \leq \lim \sup_N \lambda_{\max}(\mathbf{R})<  C_{+}.\]
\end{enumerate}
 \end{myassu}

 Under Assumption 1, $\{\mathbf{x}_l\}$ are i.i.d. and as $N\to \infty$, $\phi_l(\alpha)$ approaches its expectation according to the RMT \cite{tulino2004random}.  
Meanwhile, as $L\to \infty$, the arithmetic mean approaches the statistical expectation. 
Thus, we replace the $l$-dependent term $\phi_l(\alpha)$ by their arithmetical average $\varphi(\alpha)$ and consider
\begin{equation}
    \begin{split}
\mathcal{J}_{\mathrm{AE}} (\alpha)=\frac{1}{\left(1-\frac{N}{L}\varphi(\alpha)\right)^2}\frac{\mathbf{s}^\mathrm{H} \widetilde{\mathbf{R}}_{\mathrm{G}}^{-1} (\alpha) \widehat{\mathbf{R}}_{\mathrm{SCM}}    \widetilde{\mathbf{R}}_{\mathrm{G}}^{-1}(\alpha) \mathbf{s} }{ \left( \mathbf{s}^\mathrm{H} \widetilde{\mathbf{R}}_{\mathrm{G}}^{-1} (\alpha)  \mathbf{s}\right)^2}  
    \end{split}
    \label{Falpha2}
\end{equation} 
where  
\begin{equation}
    \begin{split}
    &\varphi(\alpha)=\frac{1}{L}\sum_{l=1}^L \phi_l(\alpha)\\
    &=\frac{1}{N}  \mathrm{Tr} \left(  \widehat{\mathbf{R}}_{\mathrm{SCM}} \widetilde{\mathbf{R}}_{\mathrm{G}}^{-1} (\alpha)\right)-  \frac{1}{N}   \frac{ \mathbf{s}^\mathrm{H}\widetilde{\mathbf{R}}_{\mathrm{G}}^{-1}(\alpha) \widehat{\mathbf{R}}_{\mathrm{SCM}} \widetilde{\mathbf{R}}_{\mathrm{G}}^{-1}(\alpha) \mathbf{s}
 }{  \mathbf{s}^\mathrm{H}\widetilde{\mathbf{R}}_{\mathrm{G}}^{-1}(\alpha) \mathbf{s}} .
 \end{split}
\end{equation}  
Furthermore, according to the Rayleigh-Ritz theorem, we have
\[\lambda_{G,\min}<\frac{ \mathbf{s}^\mathrm{H}\widetilde{\mathbf{R}}_{\mathrm{G}}^{-1}(\alpha) \widehat{\mathbf{R}}_{\mathrm{SCM}} \widetilde{\mathbf{R}}_{\mathrm{G}}^{-1}(\alpha) \mathbf{s}
 }{  \mathbf{s}^\mathrm{H}\widetilde{\mathbf{R}}_{\mathrm{G}}^{-1}(\alpha) \mathbf{s}}
 <\lambda_{G,\max},\]
 where $\lambda_{G,\min}$ and $\lambda_{G,\max}$ denote the minimum and maximum eigenvalue of $\widehat{\mathbf{R}}_{\mathrm{SCM}} \widetilde{\mathbf{R}}_{\mathrm{G}}^{-1} (\alpha)$, respectively.
 Recalling (\ref{RscmAI}) and $\mathbf{T}$ is positive-definite, we have $\widehat{\mathbf{R}}_{\mathrm{SCM}} \widetilde{\mathbf{R}}_{\mathrm{G}}^{-1} (\alpha)=\widehat{\mathbf{R}}_{\mathrm{SCM}} (\widehat{\mathbf{R}}_{\mathrm{SCM}} +\alpha \mathbf{T})^{-1}$, which indicates that all eigenvalues of  $\widehat{\mathbf{R}}_{\mathrm{SCM}} \widetilde{\mathbf{R}}_{\mathrm{G}}^{-1} (\alpha)$ are less than 1 \cite{8352743}. 
 Therefore, we have that 
 \[0<\lim_{N\to \infty} \frac{1}{N}\frac{ \mathbf{s}^\mathrm{H}\widetilde{\mathbf{R}}_{\mathrm{G}}^{-1}(\alpha) \widehat{\mathbf{R}}_{\mathrm{SCM}} \widetilde{\mathbf{R}}_{\mathrm{G}}^{-1}(\alpha) \mathbf{s}
 }{  \mathbf{s}^\mathrm{H}\widetilde{\mathbf{R}}_{\mathrm{G}}^{-1}(\alpha) \mathbf{s}} <\frac{1}{N} \to 0.\] 
Then according to the squeeze theorem, we obtain 
\begin{equation}
    \begin{split}
    	\varphi(\alpha)= \frac{1}{N}  \mathrm{Tr} \left(  \widehat{\mathbf{R}}_{\mathrm{SCM}} \widetilde{\mathbf{R}}_{\mathrm{G}}^{-1} (\alpha)\right),
    \end{split}
    \label{varphialpha2}
\end{equation} 
in the large $N,L$ regime. 
The resulting $\mathrm{S}^2$CM-AE estimator is summarized in Algorithm  \ref{alg_S2CMAE}. 
\begin{algorithm}[t] 
\caption{The $\mathrm{S}^2$CM-AE Estimator} 
\begin{enumerate} 
  \item Compute the optimized shrinkage parameter $\rho$ via a numerical search \[\rho_{\mathrm{AE}}^{\star} =\arg\min_{\rho \in (0,1)}\mathcal{J}_{\mathrm{AE}} \left( \alpha(\rho ) \right), \]
  where   $\mathcal{J}_{\mathrm{AE}} \left( \alpha \right)$ is given by 
  (\ref{Falpha2}), $\varphi(\alpha)$ is defined as (\ref{varphialpha2}), and $\alpha (\rho )=\frac{\rho (L-1) }{L(1-\rho )}$. 
  \item Compute the $\mathrm{S}^2$CM-AE via (\ref{RSCM0}).
\end{enumerate} 
\label{alg_S2CMAE} 
\end{algorithm}

For the identity target matrix $\mathbf T = \mathbf I$, we have the following  proposition  {that establishes the asymptotic equivalence of $\mathcal{J}_{\mathrm{CV}} (\alpha)$ and $\mathcal{J}_{\mathrm{AE}} (\alpha)$}:

\begin{mypro}
\label{almostsureequal}
As $N \to \infty,L \to \infty$ with $N/L \to c$, the proposed 
LOOCV cost function with $\mathbf T = \mathbf I$ approaches asymptotically to
$\mathcal{J}_{\mathrm{AE}} (\alpha)$ in (\ref{Falpha2}), i.e.,
\begin{equation}
    \begin{split}
\sup_{\alpha \in (0, +\infty)} \left| \mathcal{J}_{\mathrm{CV}}  (\alpha) -  \mathcal{J}_{\mathrm{AE}} (\alpha)  \right| \xrightarrow{a.s.} 0.
    \end{split}
    \label{propos2}
\end{equation} 
\end{mypro}

\textit{Proof}: See Appendix \ref{Pro2}. $\hfill\blacksquare$

If we choose $\mathbf{T}=\mathbf{I}$, then (\ref{Falpha2}) with  (\ref{varphialpha2}) coincides with  the cost function Eq. (17) of Mestre and Lagunas \cite{1561576}.
This indicates that the proposed $\mathrm{S}^2$CM-AE
 is equivalent to the method of  \cite{1561576} in the asymptotic regime for Gaussian disturbances and identity targets,  though they are obtained using different tools, i.e., LOOCV and RMT. 
As will be shown next, the proposed methods can be easily generalized for non-identity targets and non-Gaussian distributions.

\section{ {STE Cross-Validated for MVDR Beamforming: Compound Gaussian Distribution}} 

This section {studies the shrinkage Tyler's estimator (STE) of $\mathbf{R}$ of compound Gaussian disturbance, which is defined as the solution to}  
\begin{equation}
    \begin{split}
	\widehat{\mathbf{R}}_{\mathrm{CG}} (\rho)=(1-\rho) \frac{1}{L} \sum_{l=1}^{L} \frac{\mathbf{y}_l \mathbf{y}_l^\mathrm{H} }{ \frac{1}{N}  \mathbf{y}_l^\mathrm{H} \widehat{\mathbf{R}}_{\mathrm{CG}}^{-1} (\rho) \mathbf{y}_l }+\rho \mathbf{T}.
    \end{split}
    \label{shrinkageMest}
\end{equation} 
 As analyzed in \cite{6879466}, a solution to (\ref{shrinkageMest}) exists for a positive-definite target when $\rho\in \mathcal{S}\triangleq (\max(0,1-\frac{L}{N}),1)$. 
We focus on the choice of the shrinkage factor for MVDR beamforming in this paper. 
The  {associated} iterative process for solving (\ref{shrinkageMest}) is
\begin{equation}
    \begin{split}
	\widehat{\mathbf{R}}_{\mathrm{CG},k+1}(\rho)	= (1-\rho )\frac{1}{L} \sum_{l=1}^{L} \frac{\mathbf{y}_l \mathbf{y}_l^\mathrm{H} }{ \frac{1}{N}  \mathbf{y}_l^\mathrm{H} \widehat{\mathbf{R}}_{\mathrm{CG},k}^{-1} (\rho) \mathbf{y}_l }  +\rho \mathbf{T}, 
    \end{split}
    \label{tildeRk1}
\end{equation} 
where $\widehat{\mathbf{R}}_{\mathrm{CG},k}(\rho)$ denotes the estimate of $\mathbf R$ at the $k$th iteration. 
If the solution exists, such an iteration process converges to a static point for any positive-definite Hermitian initial matrix \cite{6913007}.   
Denote by $\widehat{\mathbf{R}}_{\mathrm{CG}, k}$ the estimate of $\mathbf{R}$  at the $k$-th iteration. 
Define a distance measure
\begin{equation}
    \begin{split}
\mathcal{D}^2(\widehat{\mathbf{R}}_{\mathrm{CG},k+1},\widehat{\mathbf{R}}_{\mathrm{CG},k})=\frac{\left\Vert  \widehat{\mathbf{R}}_{\mathrm{CG},k+1}  -\widehat{\mathbf{R}}_{\mathrm{CG},k}   \right\Vert_F}{\left\Vert  \widehat{\mathbf{R}}_{\mathrm{CG},k}  \right\Vert_F}. 
\end{split}
  \label{DistanceDefine}
\end{equation} 
A convergence criterion can then be set as
\begin{equation}
    \begin{split}
\mathcal{D}^2(\widehat{\mathbf{R}}_{\mathrm{CG},k+1},\widehat{\mathbf{R}}_{\mathrm{CG},k})<\delta,
\end{split}
  \label{stop}
\end{equation} 
 {where $\delta$ denotes the threshold. }
 {The computational complexity is $\mathcal{O}(N_{it}(N^3+L N^2))$ if $N_{it}$ iterations are used.}  
The $k$th iteration of the STE has a form similar to SCM based estimator except that the training samples $\mathbf{y}_l$ are adaptively weighted by ${1}/{ \sqrt{\frac{1}{N}  \mathbf{y}_l^\mathrm{H} \widehat{\mathbf{R}}_{\mathrm{CG},k}^{-1} \mathbf{y}_l }}$.
 
We define the STE-oracle choice of the shrinkage factor as  
\begin{equation}
    \begin{split}
    \rho_{\mathrm{STE-oracle}}=\min_{\rho} \mathcal{J} \left( \widehat{\mathbf{R}}_{\mathrm{CG}} \left(  \rho \right)   \right) ,\end{split}
    \label{STEoracle}
\end{equation} 
where $\mathcal{J}(\cdot)$ is defined in (\ref{cost00}) and $\widehat{\mathbf{R}}_{\mathrm{CG}} \left(  \rho \right) $ denotes the solution to  (\ref{shrinkageMest}). 
In this section, we also apply the LOOCV strategy to approximate $  \rho_{\mathrm{STE-oracle}}$. 
Several different implementations are introduced next.

\subsection{ {STE-CV-I}}

If  $\mathbf R$ is known, we can then define a set of fictional training samples as 
\begin{equation}
\label{qlexp} 
\mathbf{q}_l =\frac{\mathbf{y}_l  }{\sqrt{\frac{1}{N}\mathbf{y}_l^\mathrm{H} \mathbf{R}^{-1} \mathbf{y}_l}}, l=1,2,\cdots, L. 
\end{equation}  
 Recalling (\ref{ExpTyler}), we have $ \mathbb E (\mathbf{q}_l \mathbf{q}_l^H) = \mathbf R.$ Here we treat $\mathbf{R}$ as known, and then the expectation is performed on $\mathbf{y}_l$. This indicates that the SCM of $\mathbf q_l$ provides an unbiased estimate of $\mathbf R$.   
Furthermore, $\frac{ | \mathbf{s}^\mathrm{H}    \mathbf{\Sigma}^{-1}     \mathbf{q}_l|^2} { \left(\mathbf{s}^\mathrm{H}   \mathbf{\Sigma}^{-1} \mathbf{s}\right)^2}$ provides an unbiased estimate of the MVDR beamformer performance metric (\ref{cost00}) when an estimate $\mathbf \Sigma$ of $\mathbf R$ is known, i.e., 
$
\mathbb E \left( \frac{| \mathbf{s}^\mathrm{H}    \mathbf{\Sigma}^{-1}     \mathbf{q}_l|^2} { \left(\mathbf{s}^\mathrm{H}   \mathbf{\Sigma}^{-1} \mathbf{s}\right)^2}  \right) = \mathcal{J}\left( \mathbf{\Sigma} \right),  
$
 where $\mathbf{\Sigma}$ is treated as an argument and the expectation is only taken on $\mathbf{q}_l$. 
In the following, we use the one-step estimator  
\[ \widehat{\mathbf{R}}_{o}(\rho)=(1-\rho) \frac{1}{L} \sum_{l=1}^{L}  {\mathbf{q}_l \mathbf{q}_l^\mathrm{H}} +\rho \mathbf{T} 
 \]
as a proxy for $\widehat{\mathbf{R}}_{\mathrm{CG}} (\rho)$ to compute an estimate of the MVDR  beamformer performance metric. 
Similarly, $\mathcal{Q}=\{ \mathbf{q}_{l}\}_{l=1}^L$ are repeatedly split into two sets. 
For the $l$th split, the components in the set $\mathcal{Q}_l$ (with the $l$th component $\mathbf{q}_{l}$ omitted from $\mathcal{Q}$) are used for producing a covariance matrix estimate
$\widehat{\mathbf{R}}_{o,l}(\rho)=(1-\rho) \frac{1}{L-1} \sum_{i\neq l}  {\mathbf{q}_i \mathbf{q}_i^\mathrm{H}}  +\rho \mathbf{T},
$
and the remaining  {sample} $\mathbf{q}_{l}$ is spared for parameter validation.  
Substituting $\mathbf{\Sigma}$ and $\mathbf{R} $ with $\widehat{\mathbf{R}}_{o,l}(\rho)$ and $\mathbf{q}_l \mathbf{q}_l^\mathrm{H}$
in (\ref{cost00}), we obtain the LOOCV output power for the $l$th split as
 \begin{equation}
    \begin{split}
P_l
&=\frac{\mathbf{s}^\mathrm{H}    \widehat{\mathbf{R}}_{o,l}^{-1}(\rho)    \mathbf{q}_l \mathbf{q}_l^\mathrm{H} \widehat{\mathbf{R}}_{o,l}^{-1}(\rho)   \mathbf{s}}{ \left(\mathbf{s}^\mathrm{H}   \widehat{\mathbf{R}}_{o,l}^{-1}(\rho)   \mathbf{s}\right)^2}=\frac{\mathbf{s}^\mathrm{H}    \widetilde{\mathbf{R}}_{o,l}^{-1}(\alpha)    \mathbf{q}_l \mathbf{q}_l^\mathrm{H} \widetilde{\mathbf{R}}_{o,l}^{-1}(\alpha)   \mathbf{s}}{ \left(\mathbf{s}^\mathrm{H}   \widetilde{\mathbf{R}}_{o,l}^{-1}(\alpha)   \mathbf{s}\right)^2} 
    \end{split}
\end{equation} 
as a proxy for (\ref{cost00}), where
$   \widetilde{\mathbf{R}}_{o,l}(\alpha) 
=  \frac{1}{L} \sum_{i\neq l} \mathbf{q}_i \mathbf{q}_i^\mathrm{H}+\alpha \mathbf{T}$
with $\alpha=\frac{\rho (L-1)}{L(1-\rho)}$. 
Define \[\widetilde{\mathbf{R}}_{o}(\alpha)=\frac{1}{L} \sum_{ l=1}^L \frac{\mathbf{y}_l \mathbf{y}_l^\mathrm{H}}{ \frac{1}{N}  \mathbf{y}_l^\mathrm{H} \mathbf{R}^{-1} \mathbf{y}_l }+\alpha \mathbf{T} = \frac{1}{L} \sum_{l=1}^L \mathbf{q}_l \mathbf{q}_l^\mathrm{H}+\alpha \mathbf{T}, \] we then have
\[\widetilde{\mathbf{R}}_{o,l}(\alpha)
=\widetilde{\mathbf{R}}_{o}(\alpha)-\frac{1}{L}\mathbf{q}_l \mathbf{q}_l^\mathrm{H}.
\]
Similarly to the derivation from (\ref{Rglalphainv}) to (\ref{wk1lHul0}) in Sec. \ref{SecChoiceofG}, we obtain a LOOCV cost function for the compound Gaussian case by substituting $\widetilde{\mathbf{R}}_{\mathrm{G},l} (\alpha)$, $\widetilde{\mathbf{R}}_{\mathrm{G}} (\alpha)$ and $\mathbf{x}_l \mathbf{x}_l^\mathrm{H}$ by $\widetilde{\mathbf{R}}_{o,l}(\alpha)$, $\widetilde{\mathbf{R}}_{o}(\alpha)$ and $\mathbf{q}_l \mathbf{q}_l^\mathrm{H}$ respectively, then

\begin{equation}
    \begin{split}
& \mathcal{J}_{\mathrm{CV\text{-}I}} \left( \alpha | \mathbf{R} \right)  
 = \frac{1}{L} \sum_{l=1}^L  \frac{ \mathbf{s}^\mathrm{H} \widetilde{\mathbf{R}}_{o}^{-1}\left( \alpha\right) \mathbf{q}_l\mathbf{q}_{l}^\mathrm{H}   \widetilde{\mathbf{R}}_{o}^{-1}\left( \alpha \right) \mathbf{s} }{g_{o,l}^2(\alpha) \left(\mathbf{s}^\mathrm{H}   \widetilde{\mathbf{R}}_{o}^{-1}\left( \alpha \right)  \mathbf{s} \right)^2 }\\
&=\frac{N}{L} \sum_{l=1}^L  \frac{ \mathbf{s}^\mathrm{H} \widetilde{\mathbf{R}}_{o}^{-1}\left( \alpha \right) \mathbf{y}_l\mathbf{y}_l^\mathrm{H}   \widetilde{\mathbf{R}}_{o}^{-1}\left( \alpha \right) \mathbf{s} }{
   g_{o,l}^2(\alpha)
 \mathbf{y}_l^\mathrm{H} \mathbf{R}^{-1} \mathbf{y}_l
    \left(\mathbf{s}^\mathrm{H}   \widetilde{\mathbf{R}}_{o}^{-1}\left( \alpha \right)  \mathbf{s} \right)^2 },
    \end{split}
    \label{JCVI}
\end{equation}
where 
 \begin{equation}
    \begin{split}\nonumber
    	&g_{o,l}(\alpha)\\
    	&=1-\frac{1}{L} \left(\frac{\mathbf{y}_l^\mathrm{H} \widetilde{\mathbf{R}}_{o}^{-1}(\alpha) \mathbf{y}_l}{ \frac{1}{N} \mathbf{y}_l^\mathrm{H} \mathbf{R}^{-1} \mathbf{y}_l} -\frac{ \left|\mathbf{s}^\mathrm{H}\widetilde{\mathbf{R}}_{o}^{-1}(\alpha) \mathbf{y}_l \right|^2
 }{ \frac{1}{N} \mathbf{y}_l^\mathrm{H} \mathbf{R}^{-1} \mathbf{y}_l  \mathbf{s}^\mathrm{H}\widetilde{\mathbf{R}}_{o}^{-1}(\alpha) \mathbf{s}}  \right).
    \end{split}
\end{equation} 
Note that the  LOOCV function given here depends on the true $\mathbf{R}$ because  $\mathbf{R}$ is involved in $\widetilde{\mathbf{R}}_{o}(\alpha)$ and $\mathbf{q}_l$. Then similarly to \cite{6913007,5743027,ollila2020mestimators}, we
propose to substitute it by a fixed, trace-normalized, data-dependent and easy-to-compute estimate $\widehat{\mathbf{R}}$. 
Substituting $\widetilde{\mathbf{R}}_{\mathrm{G}} (\alpha)$ by $\widetilde{\mathbf{R}}_{o}(\alpha)$, choosing $\mathbf{T}=\mathbf{I}$, then similarly to the derivation from (\ref{Rglalphainv}) to (\ref{wk1lHul0}) in Sec. \ref{SecChoiceofG}, we can also obtain efficient implementation for STE-CV-I using eigenvalue decomposition.
The STE-CV-I estimator is summarized in Algorithm \ref{alg_STECVi}. 

Compared with $\mathrm{S}^2$CM-CV,  the computational cost for obtaining the shrinkage factor with STE-CV-I is about  
$\mathcal{O}(( N_G+1) (N^3+L N^2))$, due to  the involvement of an extra inverse of $\widehat{\mathbf{R}}$. Thus the total complexity of STE-CV-I is about  $\mathcal{O}(( N_G+1) (N^3+L N^2)+N_{it} (N^3+LN^2))$. For identity target, the complexity is about $\mathcal{O}(2 N^3+ {2 L N^2 + N_G LN}+N_{it} (N^3+LN^2))$.

\begin{algorithm}[t] 
\caption{The STE-CV-I Estimator} 
\textbf{Input:} A preliminary estimator $\widehat{\mathbf{R}}$.
\begin{enumerate} 
  \item Compute the optimized shrinkage parameter via a numerical search 
  \[
\rho_{\mathrm{CV\text{-}I}}^{\star} =\mathop{\mathrm{arg\, min}}\limits_{ \rho \in \mathcal{S}}\mathcal{J}_{\mathrm{CV\text{-}I}} \left( \alpha(\rho )| \widehat{\mathbf{R}} \right), \]
  where $\mathcal{J}_{\mathrm{CV\text{-}I}} \left( \alpha \right)$ is defined as (\ref{JCVI})  and $\alpha (\rho )=\frac{\rho (L-1) }{L(1-\rho )}$.
  \item {Compute iteratively} the {unique} solution to 
  \[\widehat{\mathbf{R}}_{\mathrm{CV\text{-}I}}=(1-\rho_{\mathrm{CV\text{-}I}}^{\star} ) \frac{1}{L} \sum_{l=1}^{L} \frac{\mathbf{y}_l \mathbf{y}_l^\mathrm{H} }{ \frac{1}{N}  \mathbf{y}_l^\mathrm{H} \widehat{\mathbf{R}}_{\mathrm{CV\text{-}I}}^{-1} \mathbf{y}_l }+\rho_{\mathrm{CV\text{-}I}}^{\star}  \mathbf{T}. \]
\end{enumerate} 
\label{alg_STECVi} 
\end{algorithm}

\subsection{STE-CV-II}
With STE-CV-I, a fixed shrinkage factor is used through the STE iterations. The performance depends on the shrinkage factor, which in turn depends on the initial plug-in CM $\widehat{\mathbf{R}}$. An inaccurate substitution CM $\widehat{\mathbf{R}}$ may limit the performance. Alternatively, we can also update $\widehat{\mathbf{R}}$ as the refined CM estimate generated during the STE iterations. This leads to adaptive update of the shrinkage factor, referred to as STE-CV-II hereafter.

Define the normalized samples at the $k$th iteration as
\[
\mathbf{q}_{k,l} =\frac{\mathbf{y}_l }{ \sqrt{\frac{1}{N}  \mathbf{y}_l^\mathrm{H} \widehat{\mathbf{R}}_{\mathrm{CG},k}^{-1} \mathbf{y}_l }}.
\]
Then 
the {STE-CV-II} estimate at the $k$th iteration  
\begin{equation}
    \begin{split}
\widehat{\mathbf{R}}_{\mathrm{CG},k+1} (\rho)= (1-\rho )\frac{1}{L} \sum_{l=1}^{L} \frac{\mathbf{y}_{l}  \mathbf{y}_{l}^\mathrm{H}}{\frac{1}{N}  \mathbf{y}_l^\mathrm{H} \widehat{\mathbf{R}}_{\mathrm{CG},k}^{-1}  \mathbf{y}_l} +\rho \mathbf{T},
  \end{split}
  \label{RCGK1}
\end{equation} 
where $\widehat{\mathbf{R}}_{\mathrm{CG},k}$ denotes the initial estimate of $\mathbf{R}$ at the $k$th iteration,  
  can be rewritten as
\begin{equation}
    \begin{split} 
    { \widehat{\mathbf{R}}_{\mathrm{CG},k+1}}	 (\rho)= (1-\rho )\frac{1}{L} \sum_{l=1}^{L} \mathbf{q}_{k,l}  \mathbf{q}_{k,l} ^\mathrm{H} +\rho \mathbf{T}.
  \end{split}
  \label{RCGK2}
\end{equation} 
With (\ref{RCGK2}), we treat the current estimate $\widehat{\mathbf{R}}_{\mathrm{CG}, k}$ of $\mathbf{R}$ as constant.  
Clearly, (\ref{RCGK2}) resembles  (\ref{RSCM0}). 
By similar derivation in Sec. \ref{SecChoiceofG}, we obtain the LOOCV cost function  for STE-CV-II as
\begin{equation}
    \begin{split}
&\mathcal{J}_{\mathrm{CV\text{-}II},k} \left( \alpha\right)\\
&=\frac{N}{L} \sum_{l=1}^L \frac{1}{  g_{\mathrm{CG},k,l}^2(\alpha)} \cdot \frac{ \mathbf{s}^\mathrm{H} \widetilde{\mathbf{R}}_{\mathrm{CG},k}^{-1}\left( \alpha \right) \mathbf{y}_l\mathbf{y}_l^\mathrm{H}   \widetilde{\mathbf{R}}_{\mathrm{CG},k}^{-1}\left( \alpha \right) \mathbf{s} }{
 \mathbf{y}_l^\mathrm{H} \widehat{\mathbf{R}}_{\mathrm{CG},k}^{-1} \mathbf{y}_l
    \left(\mathbf{s}^\mathrm{H}   \widetilde{\mathbf{R}}_{\mathrm{CG},k}^{-1}\left( \alpha \right)  \mathbf{s} \right)^2 },
    \end{split}
    \label{JCVII}
\end{equation}
where
\begin{equation}
    \begin{split}
\widetilde{\mathbf{R}}_{\mathrm{CG},k} (  \alpha )	=\frac{1}{L} \sum_{l=1}^L \frac{\mathbf{y}_l \mathbf{y}_l^\mathrm{H} }{ \frac{1}{N} \mathbf{y}_l^\mathrm{H} \widehat{\mathbf{R}}_{\mathrm{CG},k}^{-1} \mathbf{y}_l}+\alpha \mathbf{T},
   \end{split}
   \label{RCGKalphaK}
\end{equation}
and 
 \begin{equation}
    \begin{split}\nonumber
    	&g_{\mathrm{CG},k,l}(\alpha)\\
    	&=1-\frac{N}{L} \left(\frac{\mathbf{y}_l^\mathrm{H} \widetilde{\mathbf{R}}_{\mathrm{CG},k}^{-1}(\alpha) \mathbf{y}_l}{ \mathbf{y}_l^\mathrm{H}   \widehat{\mathbf{R}}_{\mathrm{CG},k}^{-1}    \mathbf{y}_l} -\frac{ \left|\mathbf{s}^\mathrm{H}\widetilde{\mathbf{R}}_{\mathrm{CG},k}^{-1}(\alpha) \mathbf{y}_l \right|^2
 }{  \mathbf{y}_l^\mathrm{H} \widehat{\mathbf{R}}_{\mathrm{CG},k}^{-1} \mathbf{y}_l  \mathbf{s}^\mathrm{H}\widetilde{\mathbf{R}}_{\mathrm{CG},k}^{-1}(\alpha) \mathbf{s}}  \right).
    \end{split}
\end{equation} 
The resulting STE-CV-II estimator is summarized in Algorithm \ref{alg_STECVii}.
\begin{algorithm}[t] 
\caption{The STE-CV-II Estimator} 
\textbf{Input:} Initial point $\widehat{\mathbf{R}}_{\mathrm{CG},0}$.\\
\textbf{Initialize:} $k=0$.\\
\textbf{Repeat:}
\begin{enumerate} 
  \item Compute the optimized shrinkage parameter $\rho$ via a numerical search 
\[\rho_{\mathrm{CV\text{-}II},k}^{\star} =\mathop{\mathrm{arg\, min}}\limits_{\rho  \in \mathcal{S}}\mathcal{J}_{\mathrm{CV\text{-}II},k} \left( \alpha(\rho ) \right),\]
  where $\mathcal{J}_{\mathrm{CV\text{-}II},k} \left( \alpha \right)$ is defined as (\ref{JCVII})   and $\alpha (\rho )=\frac{\rho (L-1) }{L(1-\rho )}$.
  \item Update $\widehat{\mathbf{R}}_{\mathrm{CG},k+1}=\widehat{\mathbf{R}}_{\mathrm{CG},k+1}(\rho_{\mathrm{CV\text{-}II},k}^{\star})$ via (\ref{RCGK1}).
  \item $k\gets k+1$.
\end{enumerate} 
  \textbf{Until:} The termination criterion (\ref{stop}) is met.
\label{alg_STECVii} 
\end{algorithm}

\subsection{STE-AE}
We can also extend the $\mathrm{S}^2$CM-AE in  {Section III-B to} the compound Gaussian case, yielding {STE-AE}. Here we use the similar process as {STE-CV-II} and update the shrinkage factor at each iteration. 
The resultant estimator at the $k$th iteration is the same as (\ref{RCGK1}). By substituting $\mathbf{y}_l$ by $\mathbf{q}_{k,l}$ in (\ref{Falpha2}), we obtain the cost function, i.e.,
\begin{equation}
    \begin{split}
&\mathcal{J}_{\mathrm{AE},k} (\alpha)\\
&=\frac{1}{h_{\mathrm{AE},k}^2(\alpha)} \cdot \frac{N}{L} \sum_{l=1}^L \frac{ \mathbf{s}^\mathrm{H} \widetilde{\mathbf{R}}_{\mathrm{CG},k}^{-1}\left( \alpha \right) \mathbf{y}_l\mathbf{y}_l^\mathrm{H}   \widetilde{\mathbf{R}}_{\mathrm{CG},k}^{-1}\left( \alpha \right) \mathbf{s} }{
 \mathbf{y}_l^\mathrm{H} \widehat{\mathbf{R}}_{\mathrm{CG},k}^{-1} \mathbf{y}_l
    \left(\mathbf{s}^\mathrm{H}   \widetilde{\mathbf{R}}_{\mathrm{CG},k}^{-1}\left( \alpha \right)  \mathbf{s} \right)^2 },
    \end{split}
    \label{JAEK1}
\end{equation} 
where
 \begin{equation}
    \begin{split}\nonumber
    	h_{\mathrm{AE},k}(\alpha)=1-\frac{1}{L} \left(   \frac{1}{L} \sum_{l=1}^L      \frac{\mathbf{y}_l^\mathrm{H} \widetilde{\mathbf{R}}_{\mathrm{CG},k}^{-1}(\alpha) \mathbf{y}_l}{ \frac{1}{N} \mathbf{y}_l^\mathrm{H}   \widehat{\mathbf{R}}_{\mathrm{CG},k}^{-1}    \mathbf{y}_l}  \right),
    \end{split}
\end{equation} 
and $\widetilde{\mathbf{R}}_{\mathrm{CG},k} (\alpha)$ is defined in (\ref{RCGKalphaK}). 
The STE-AE estimator is summarized in Algorithm \ref{alg_STEAE}.  {Its computational cost for obtaining the shrinkage factor is about $ {\mathcal{O}(  N_G  (N^3+L N^2))}$ and $ {\mathcal{O}(2 N^3+2 L N^2 + N_G LN)}$ with non-identity and identity target, respectively. Since we update the shrinkage factor at every iteration, the corresponding  overall complexity is about $ {\mathcal{O}(N_{it} N_G (N^3+L N^2))}$ and $ {\mathcal{O}(N_{it}(3 N^3+3 L N^2 + N_G LN))}$. 
The STE-CV-II method has the same order of complexity as the STE-AE}.
\begin{algorithm}[t] 
\caption{The STE-AE Estimator} 
\textbf{Input:} Initial point $\widehat{\mathbf{R}}_{\mathrm{CG},0}$.\\
\textbf{Initialize:} $k=0$.\\
\textbf{Repeat:}
\begin{enumerate} 
  \item Compute the optimized shrinkage factor $\rho$ via a numerical search 
\[\rho_{\mathrm{AE},k}^{\star} =\mathop{\mathrm{arg\, min}}\limits_{\rho\in \mathcal{S}}\mathcal{J}_{\mathrm{AE},k} \left( \alpha(\rho) \right),\]
  where $\mathcal{J}_{\mathrm{AE},k} \left( \alpha \right)$ is defined as (\ref{JAEK1}) and $\alpha (\rho )=\frac{\rho (L-1) }{L(1-\rho )}$.
  \item  Update $\widehat{\mathbf{R}}_{\mathrm{CG},k+1}=\widehat{\mathbf{R}}_{\mathrm{CG},k+1}(\rho_{\mathrm{AE},k}^{\star})$ via (\ref{RCGK1}).
  \item $k\gets k+1$.
\end{enumerate} 
  \textbf{Until:}   The termination criterion (\ref{stop}) is met.
\label{alg_STEAE} 
\end{algorithm}

\subsection{Remarks}
We have several remarks in order regarding the proposed methods.

{\emph{Remark 1}: With the STE-CV-I estimator proposed for compound-Gaussian data, the LOOCV strategy is applied to the weighted samples $\{\frac{\mathbf{y}_l }{\sqrt{\frac{1}{N}\mathbf{y}_l^\mathrm{H} \widehat{\mathbf{R}}^{-1} \mathbf{y}_l}}\}$, where $\widehat{\mathbf{R}}$ may be an estimate of $\mathbf{R}$ obtained from all the samples and treated as constant while applying LOOCV.  
An alternative treatment is to split the samples $\{\mathbf y_l\}$ into two sets for $L$ times, similarly to the Gaussian case. A shrinkage CM estimate $\widehat{\mathbf R}_{\rho, l}$ is generated from the training subset only using iterations similarly to  (\ref{tildeRk1}) (with $\mathbf y_l$ omitted) until convergence and an MVDR filter {$\mathbf w_{\rho, l}=\widehat{\mathbf R}_{\rho, l}^{-1} \mathbf s/(\mathbf s^H\widehat{\mathbf R}_{\rho, l}^{-1} \mathbf s)$} is then produced for each split for a given shrinkage factor. The sum of the weighted output power {$\sum_{l=1}^L\frac{1}{{\frac{1}{N}\mathbf{y}_l^\mathrm{H} \widehat{\mathbf{R}}^{-1} \mathbf{y}_l}} \frac{|\mathbf s^H\widehat{  \mathbf R}_{\rho, l}^{-1}  \mathbf y_l|^2}{    |\mathbf s^H \widehat{\mathbf R}_{\rho, l}^{-1} \mathbf s|^2} $}  
 is then selected as the cost function, which can be minimized to yield a shrinkage factor choice. This approach, though following the LOOCV principle with strict splits of the training and validation subsets, exhibits significantly higher complexity $\mathcal{O}(LN_GN_{it}(N^3+L N^2))$ as compared to STE-CV-I, STE-CV-II and STE-AE.

{\emph{Remark 2}: 
STE-CV-II and STE-AE apply the LOOCV strategy to the samples $\mathbf{q}_{k,l}$ weighed using the current estimate $\widehat{\mathbf{R}}_{\mathrm{CG},k}$ as constant at each iteration}. The variable plug-in estimators generally lead to performance better than that of STE-CV-I which depends on the choice of the fixed plug-in estimator.

 \emph{Remark 3}: {STE-CV-I} fixes a global shrinkage factor and then obtains a solution to the  STE fixed-point equation in (\ref{shrinkageMest}) by an iterative process. The existence, uniqueness, and convergence of a solution to such STE fixed-point equations  has been studied in \cite{6879466}. Meanwhile, theoretical study of the existence of an solution and the convergence of the iterative algorithm is unavailable for  {STE-CV-II}. Later, numerical studies will show that the {STE-CV-II} choice still leads to converging solutions.

\section{Numerical Examples} 
In this section,  we present simulations under several different scenarios to validate the proposed estimators. 
Consider the MVDR weight
\begin{equation}
    \begin{split}
	\widehat{\mathbf{w}} =\frac{\widehat{\mathbf{R}}^{-1} \mathbf{s}}{ \mathbf{s}^\mathrm{H}   \widehat{\mathbf{R}}^{-1} \mathbf{s}}.
    \end{split}
\end{equation} 
The output power of the signal and disturbance are $\left|\widehat{\mathbf{w}}^\mathrm{H}\mathbf{s} \right|^2=1$ and $\widehat{\mathbf{w}}^\mathrm{H}  \mathbf{R}_c \widehat{\mathbf{w}}$, respectively. Thus the output signal to disturbance ratio (SDR) is given as
\begin{equation}
    \begin{split}
	\text{SDR}=\frac{\left|\widehat{\mathbf{w}}^\mathrm{H}\mathbf{s} \right|^2}{\widehat{\mathbf{w}}^\mathrm{H}  \mathbf{R}_c \widehat{\mathbf{w}}}=\frac{\left| \mathbf{s}^\mathrm{H} \widehat{\mathbf{R}}^{-1} \mathbf{s}  \right|^2 }{\mathbf{s}^\mathrm{H} \widehat{\mathbf{R}}^{-1} \mathbf{R}_c \widehat{\mathbf{R}}^{-1} \mathbf{s}} .
    \end{split}
\end{equation} 
The optimal SDR is achieved at $\widehat{\mathbf{R}}=\mathbf{R}$, i.e., $\text{SDR}_{opt}=\mathbf{s}^\mathrm{H}  \mathbf{R}_c^{-1} \mathbf{s}$. Recall that $\mathbf{R}_c=\mathbb{E}(\tau) \mathbf{R}$. 
Following  \cite{1263229,4101326,9052470}, we will use the SDR loss
\begin{equation}
    \begin{split}
     \mathrm{SL} (\widehat{\mathbf{R}}) \triangleq \frac{\text{SDR}}{\text{SDR}_{opt}}=\frac{\left| \mathbf{s}^\mathrm{H} \widehat{\mathbf{R}}^{-1} \mathbf{s}  \right|^2 }{\left(\mathbf{s}^\mathrm{H} \widehat{\mathbf{R}}^{-1} \mathbf{R} \widehat{\mathbf{R}}^{-1} \mathbf{s} \right)
\left(\mathbf{s}^\mathrm{H} {\mathbf{R}}^{-1} \mathbf{s} \right)} 
	\label{scnrS1}
    \end{split}
\end{equation}
to evaluate the performance of the MVDR beamformer.

\begin{figure*}[!t]
\centering
    \subfloat[]{
    \includegraphics[width=3.5in]{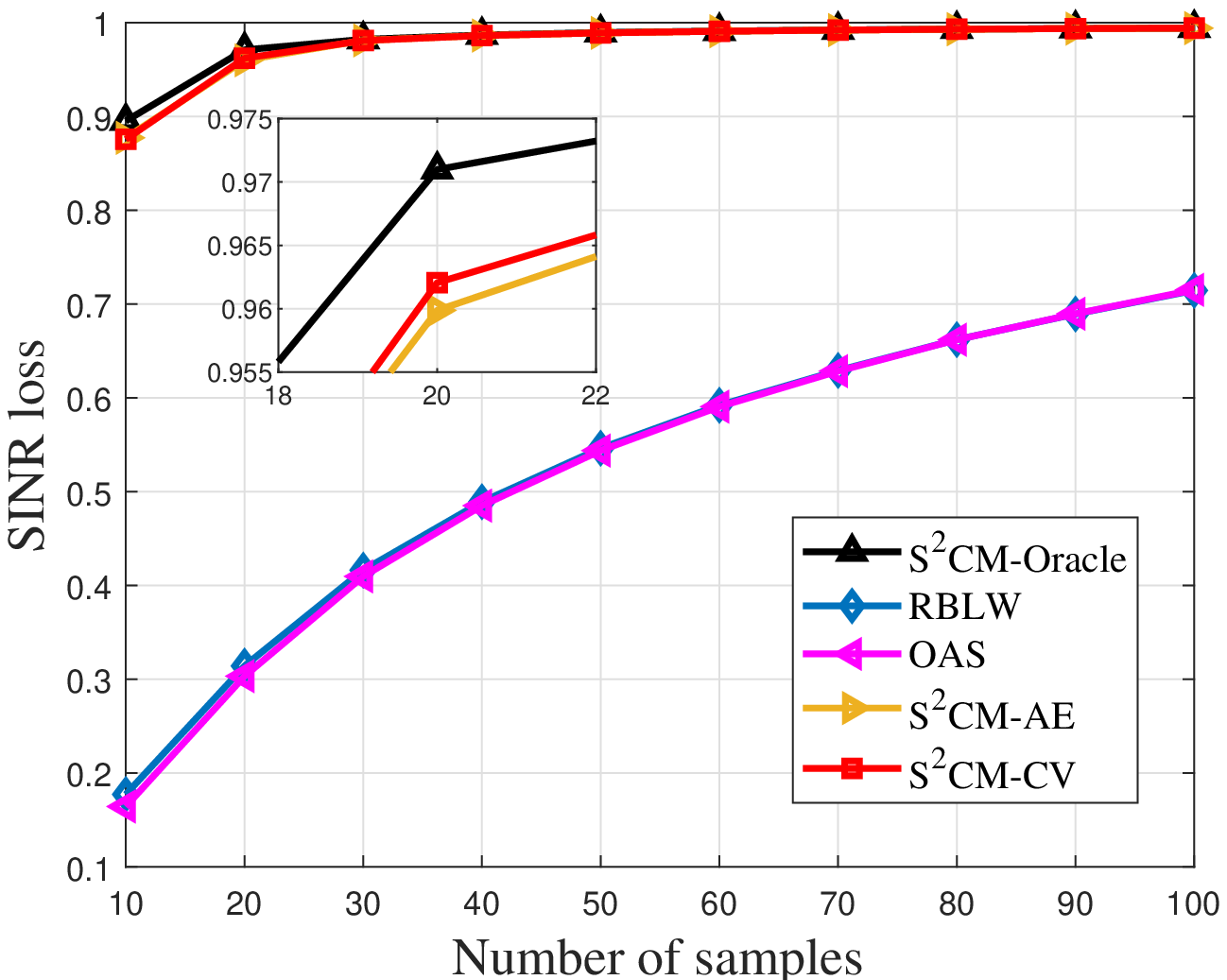}\label{fig_first_case}}\
    \subfloat[]{
    \includegraphics[width=3.5in]{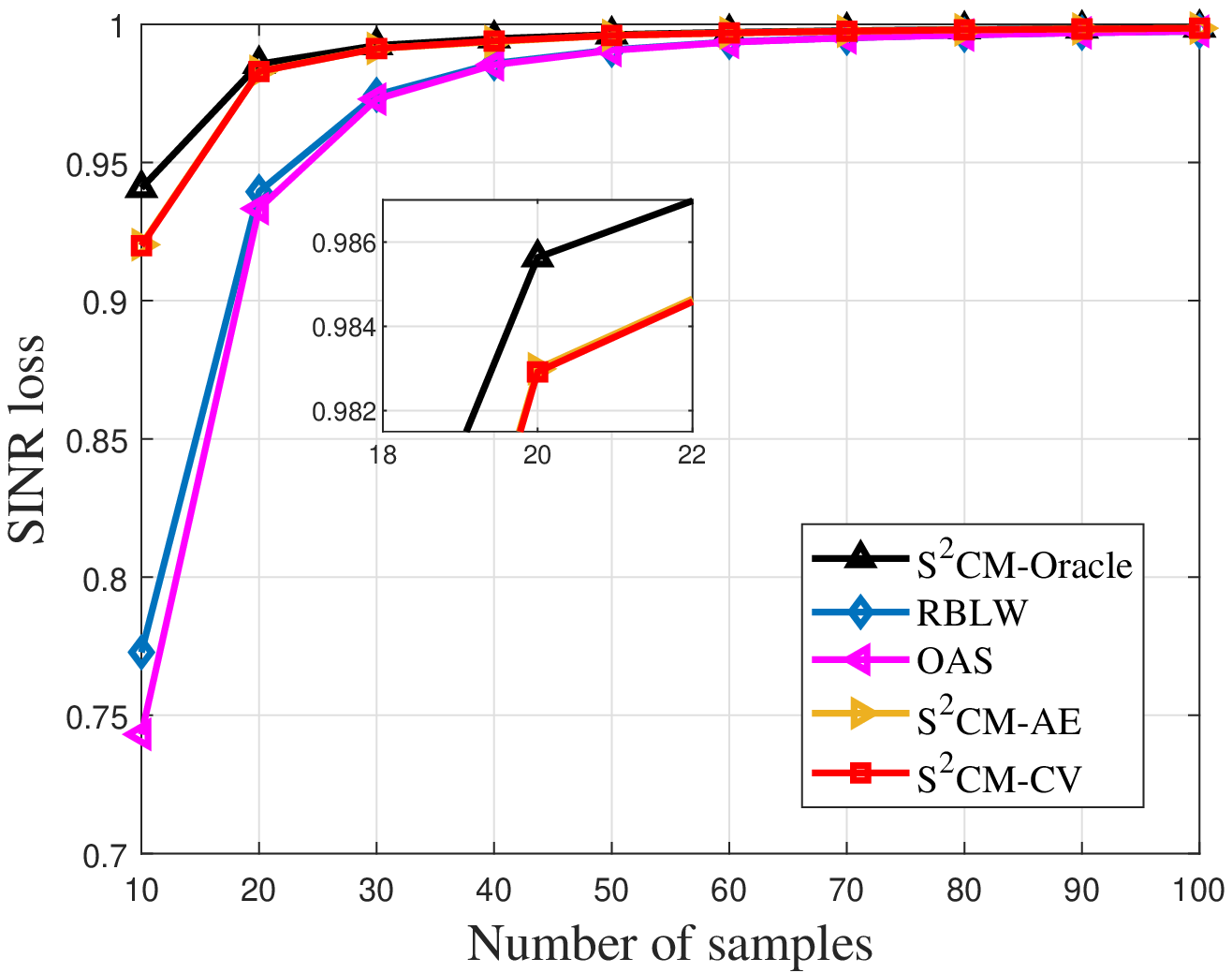}\label{fig_second_case}}\\
\caption{ SINR loss versus the number of samples $L$ with identity target. (a) $N=20$; (b) $N=40$.} 
\label{fig_G_noscenario_SL1}
\end{figure*}

\subsection{Gaussian  Samples}
\label{SimulationG}
We first consider an interference suppression case under a uniform linear array with $N$ antennas.  {Here the disturbance denotes the interference plus noise.}
The array observation is modeled as
\[  \mathbf{y}_l=\mathbf{A} \mathbf{s}_l+\mathbf{n}_l= \sum_{i=1}^Q  \sqrt{P_i} \mathbf{a}(\theta_i  ) s_{i,l}+\mathbf{n}_l, \]
where 
\[ \mathbf{A}=\left[\sqrt{P_1} \mathbf{a}(\theta_1  ), \cdots, \sqrt{P_L} \mathbf{a}(\theta_L  )\right],   \]
\[ \mathbf{a} \left( \theta_i  \right)=\left[ 1,e^{j\pi \sin (\theta_i)},\cdots,e^{j\pi \sin (\theta_i) (N-1)} \right]^{\mathrm{T}}, l=1, \cdots, L, \]
$\mathbf{s}_l=[s_{1,l},\cdots,s_{N,l}]^\mathrm{T}$ denotes the  interference waveform whose  entries are i.i.d. and follow a Gaussian distribution with zero mean and unit variance, 
$\mathbf{n}_l$ denotes the complex additive Gaussian white noise (AWGN) with mean  zero and covariance matrix $\sigma^2 \mathbf{I}$, $\theta_i, i=1, \cdots, Q$, denote the DOAs of the interferences. 
The true CM of interference-plus-noise is $\mathbf{R}=\mathbf{A} \mathbf{A}^{\mathrm{H}}+\sigma^2 \mathbf{I}$.
Here we set  $N=20$ and the DOAs of interference are from $30^\circ$ to $70^\circ$ with step $5^\circ$. The elements spacing is set as a half of wavelength.
The DOA of signal is $\theta_{tgt}=0^\circ$.
We set the interference-to-noise ratio $\mathrm{INR}= 30$ dB, i.e., $P_1=\cdots=P_Q=10^{30/10}$ when $\sigma^2=1$.

 We first consider the identity target. We compare the proposed estimator {$\mathrm{S}^2$CM-CV} with the following CM estimators designed for Gaussian samples: $\mathrm{S}^2$CM-AE\footnote{As stated in Sec. III. B, the proposed $\mathrm{S}^2$CM-AE is equivalent to the method of  \cite{1561576} for Gaussian disturbance and identity target.}, Rao-Blackwell Ledoit-Wolf (RBLW) \cite{LEDOIT2004365} and oracle approximating
shrinkage (OAS) \cite{5484583}. 
We also include the unattainable $\mathrm{S}^2$CM-oracle estimator defined in (\ref{SSCMoracle}) to show an upper bound of the $\mathrm{S}^2$CM estimator. We can classify these estimators  into two classes: the searching-based algorithms ($\mathrm{S}^2$CM-CV, $\mathrm{S}^2$CM-AE and $\mathrm{S}^2$CM-oracle) and the non-searching ones (RBLW and OAS). The two non-searching estimators aim at minimizing the MSE of estimating the CM and exhibit lower computational costs. 
\begin{figure}[!t]
\centering
\includegraphics[width=3.5in]{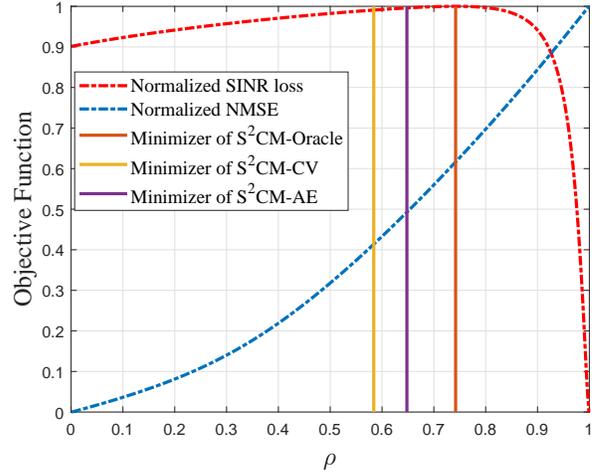}
\caption{The  normalized NMSE and SINR loss versus the shrinkage factor $\rho$ with the identity target. }
\label{fig_sim}
\end{figure}
Fig. \ref{fig_G_noscenario_SL1} shows the SIR loss with $10\le L\le 100$. 
For each abscissa, 2000 Monte-Carlo experiments are performed. 
We can see that OAS and RBLW achieve similar performance. 
The proposed $\mathrm{S}^2$CM-CV estimator has slightly better performance than $\mathrm{S}^2$CM-AE. 
They can both outperform other non-searching estimators and are close to the oracle estimator.

\begin{figure*}[!t]
\centering
    \subfloat[]{
    \includegraphics[width=3.5in]{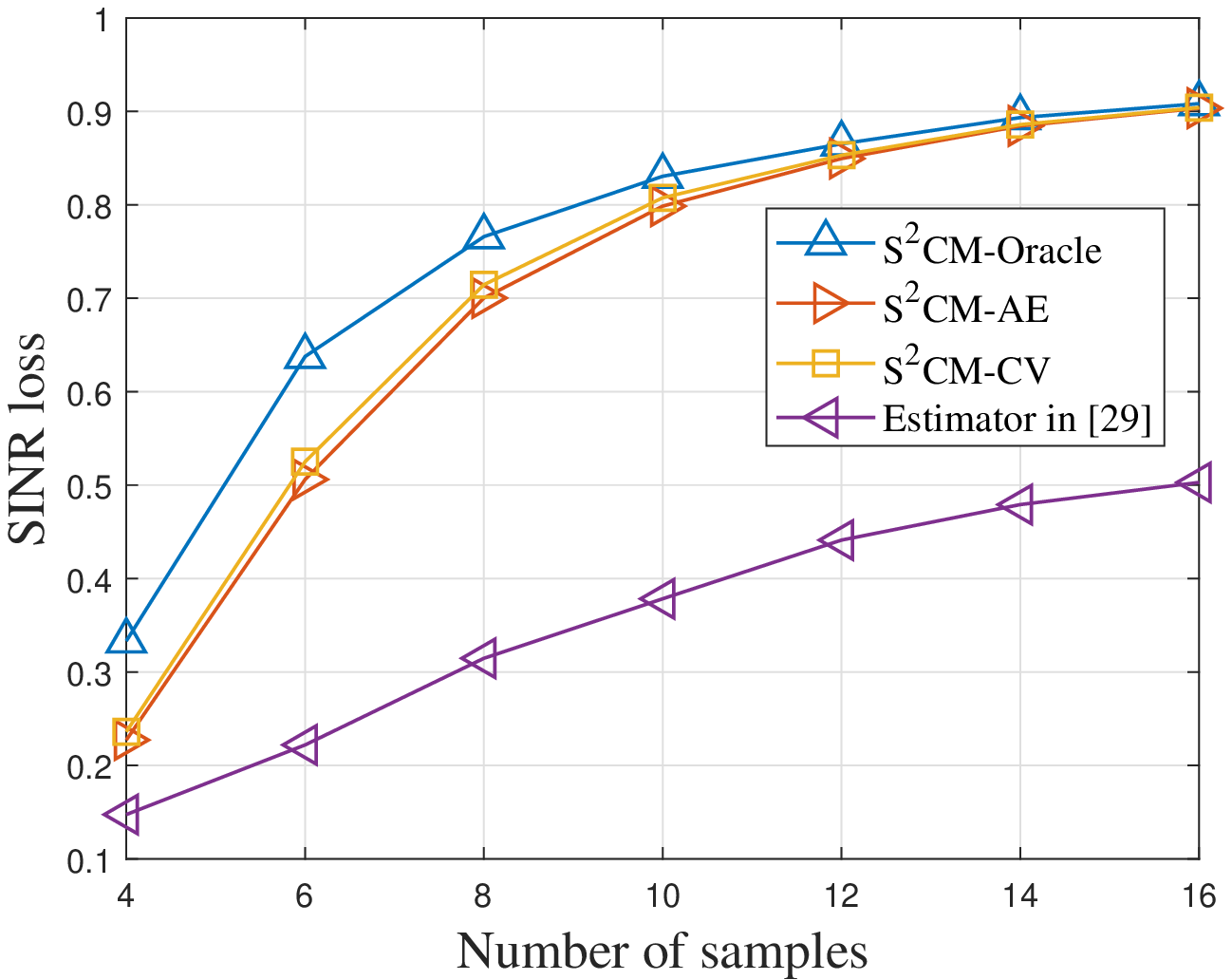}\label{fig_ni_first_case}}\
    \subfloat[]{
    \includegraphics[width=3.5in]{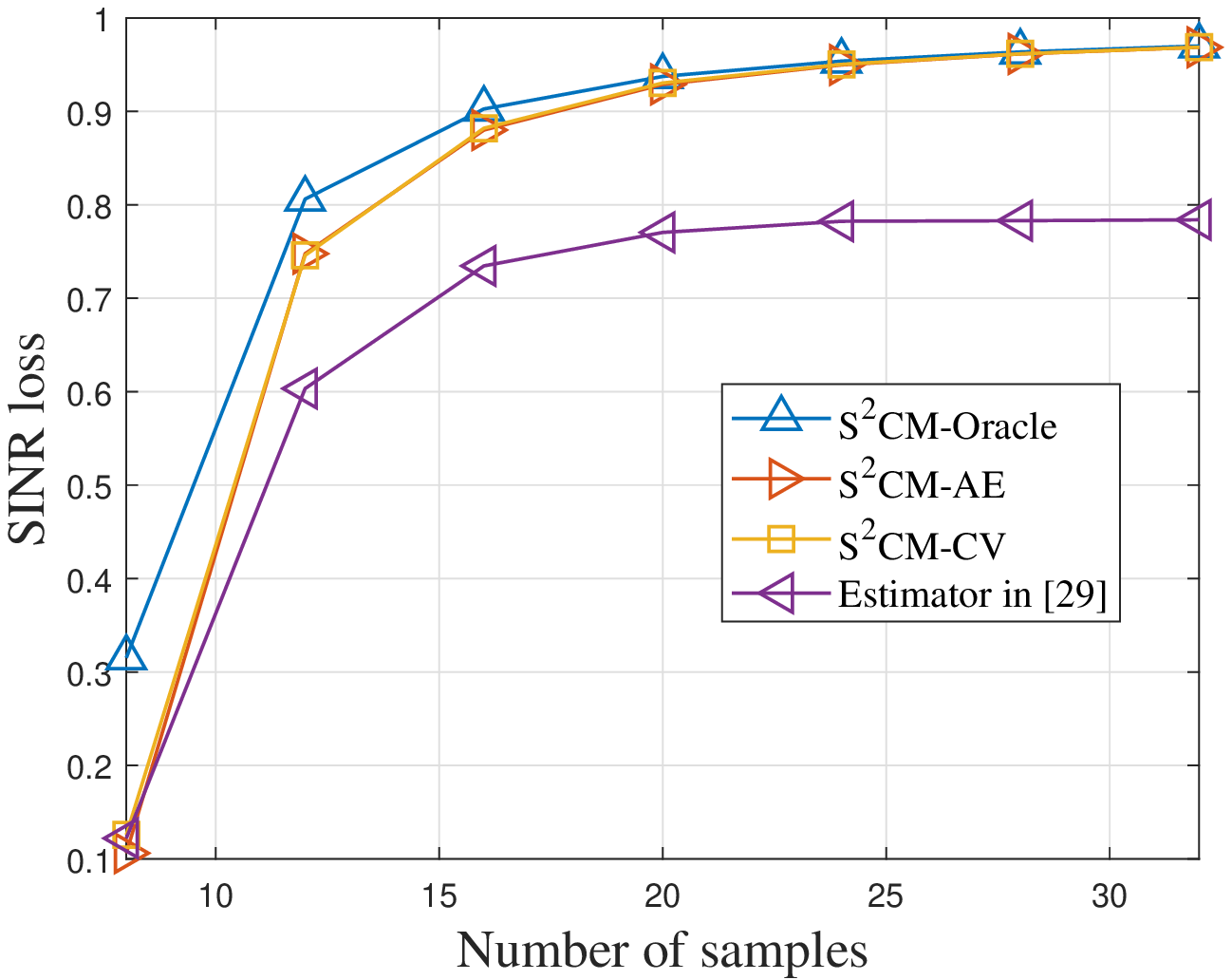}\label{fig_ni_second_case}}\\
\caption{SINR loss versus the number of samples $L$ for Gaussian interference and non-identity target. (a) $N=8$; (b) $N=16$.}
\label{fig_G_noscenario_SL0}
\end{figure*}

Moreover, we consider the normalized MSE  (NMSE) for CM estimation, i.e.,
 \[  \mathrm{NMSE}=\mathbb{E}  \left(   \frac{\left\Vert {\mathbf{R}}/{\mathrm{Tr}(  \mathbf{R} )}   -{\widehat{\mathbf{R}}}/{\mathrm{Tr}(  \widehat{\mathbf{R}} )}     \right\Vert_F^2 }{\left\Vert {\mathbf{R}}/{\mathrm{Tr}(  \mathbf{R} )}   \right\Vert_F^2}   \right).  \]
For ease of comparison, we affine the range of NMSE and SIR loss on [0,1] by $\bar{x}=(x-x_{\min})/(x_{\max}-x_{\min})$, which are referred as normalized NMSE and SIR loss,  respectively.
Fig. \ref{fig_sim}  shows the normalized NMSE and SIR loss versus the shrinkage factor $\rho$. Here we fix $L=10$, $N=20$ and other parameters are same as before. For more intuitive comparison, we also include the shrinkage factor of the oracle estimator, $\mathrm{S}^2$CM-CV and $\mathrm{S}^2$CM-AE. 
From Fig. \ref{fig_sim}, we can see that the shrinkage factor chosen by $\mathrm{S}^2$CM-CV is close to that of $\mathrm{S}^2$CM-AE, which agrees with the analysis in \emph{Proposition} \ref{almostsureequal}.
Though $\mathrm{S}^2$CM-CV and $\mathrm{S}^2$CM-AE can achieve near-optimal SIR loss performance, they do not minimize the NMSE.
This is because all of our proposed algorithms in this paper are aimed to optimize the MVDR beamformers. Thus they can not promise a better performance than the existing algorithms on other metrics, such as NMSE.

{Many of the existing
shrinkage methods are derived for identity shrinkage targets and it is unclear how to extend them to non-identity targets.  
By contrast, our proposed estimators are applicable to arbitrary positive-definite and well-conditioned targets.  
In order to show the performance of our proposed algorithms with  non-identity target matrices, following \cite{4490277, shang2019glrt}, we  {consider a target} $\mathbf{T}=\mathbf{R}\odot \mathbf{t}\mathbf{t}^\mathrm{T}$. Each entry of $\mathbf{t}$ follows a Gaussian distribution with unit mean and  variance $\sigma_t^2=0.1$, which is modeled as the error of prior knowledge.   We set $N=8, 16$ and other parameters are same as before, and  compare our proposed estimators with the estimator of \cite{4490277}, which can be used for the non-identity target.
From Fig. \ref{fig_G_noscenario_SL0}, we see that the proposed estimators can also work effectively with non-identity target matrices and outperform the estimator of \cite{4490277}.

\subsection{Compound Gaussian Case}
\label{SimulationCG}
In this section, we show the performance  of the robust shrinkage estimators with identity targets under the STAP application. In STAP, the target-free training samples collected from adjacent range rings are modeled as the sum of the clutter and noise:
 \begin{equation}  
    \begin{split}
	&\mathbf{y}=\sqrt{\tau}\mathbf{x}+\mathbf{n}, 
    \end{split}
\label{yls}
\end{equation}
where the AWGN $\mathbf{n} \sim \mathcal{CN}\left( \mathbf{0}, \sigma^2 \mathbf{I}_N \right)$. The texture $\tau_l$ follows a Gamma distribution \cite{249129} of shape parameter $\nu$ and scale parameter $1/\nu$, i.e., $\tau_l \sim \Gamma (\nu,1/\nu)$. Here we set $\nu=4.5$. 
The generated samples $\mathbf{y}_l$ follow a zero-mean CES distribution with $\mathbf{x}$ modeled as
\begin{equation}
    \begin{split}
   \mathbf{x}=\sum_{i=1}^{N_c} \xi_{i} \mathbf{a} \left(  f_{d,i},f_{s,i} \right) = \mathbf{A} \mathbf{e},
   \label{Xsum}
    \end{split}
\end{equation}
where  $\xi_{i}$ denotes the echo gain of the $i$-th clutter patch which follows a complex Gaussian  distribution with zero mean and unit variance, $\mathbf{A}=[\mathbf{a} \left( f_{d,1},f_{s,1} \right), \cdots, \mathbf{a} \left( f_{d,N_c},f_{s,N_c} \right)]$, $\mathbf{e}=[\xi_{1}, \cdots, \xi_{N_c}]^\mathrm{T}$, 
$N_c$ is the number of clutter patches uniformly distributed in the azimuth angle, and $\mathbf{a} \left( f_{d,i},f_{s,i} \right)$ denotes the space-time steering vector of the $i$-th clutter patch: 
\begin{equation}
    \begin{split}
   \mathbf{a} \left( f_{d,i},f_{s,i} \right)=\mathbf{a}_d \left( f_{d,i} \right) \otimes  \mathbf{a}_s \left( f_{s,i} \right)\in \mathbb{C}^{N_t N_s\times 1}, 
   \label{afdfs}
    \end{split}
\end{equation}
where
\begin{equation}
    \begin{split}
   &\mathbf{a}_d \left( f_{d,i} \right)=\left[ 1,e^{j2\pi f_{d,i}},\cdots,e^{j2\pi (N_t-1) f_{d,i}} \right]^{\mathrm{T}}\in \mathbb{C}^{N_t \times 1}\\
   &\mathbf{a}_s \left( f_{s,i} \right)=\left[ 1,e^{j2\pi f_{s,i}},\cdots,e^{j2\pi (N_s-1) f_{s,i}} \right]^{\mathrm{T}}\in \mathbb{C}^{N_s \times 1} 
   \label{adas}
    \end{split}
\end{equation}
are the temporal and spatial steering vectors, respectively, 
$f_{d,i}=\left( 2v_a/\lambda f_r \right)\cos(\phi_i) $ denotes the normalized Doppler frequency,  $f_{s,i}=\left( d/\lambda \right)\cos(\phi_i) $ denotes the normalized spatial frequency, $v_a$ is the speed of the platform and $\phi_i$ the direction of the $i$-th clutter patch with respect to the array. 
In the cell under test (CUT), the received clutter and noise is given as $\mathbf{y}_{0}=\sqrt{\tau_0} \mathbf{x}_0 +\mathbf{n}_0$, where $\mathbf{x}_0$ is modeled as (\ref{Xsum}), $\mathbf{n}_0\sim \mathcal{CN}\left( \mathbf{0}, \sigma^2 \mathbf{I}_N \right)$ and  $\tau_0$ is an unknown  constant. Then the clutter to noise ratio (CNR) in the CUT is defined as   $\mathrm{CNR} = \frac{  \tau_0  N_c  }{\sigma^2}$, which is set to 1000 (30 dB).
 The true CM in the cell under test (CUT) is computed as
\begin{equation}
    \begin{split}
   \mathbf{R}= \tau_0\sum_{i=1}^{N_c}  \mathbf{a} \left(  f_{d,i},f_{s,i}  \right)\mathbf{a}^\mathrm{H} \left(  f_{d,i},f_{s,i}\right)+ \sigma^2 \mathbf{I}_N.
   \label{Rsum}
    \end{split}
\end{equation}

Here we set the normalized Doppler frequency $f_{d,t}$ and spatial frequency $f_{s,t}$ of the target as 0.2 and 0.5, respectively. The number of elements $N_s=4$ and the number of pulses $N_t = 8$. We choose $\mathbf{T}=\mathbf{I}$. Thus we have $N=N_s N_t=32$. The interelement spacing of the ULA is $d=\lambda_0/2$, where $\lambda_0$ denotes the wavelength. Each clutter ring is divided into 401 clutter patches. Other radar parameters are set as: the carrier frequency $f_0=1.2$GHz, the PRF $f_r=2 $KHz, the platform velocity $v=125$m/s, the platform altitude $3000$m.
To show a large range of $L$ including both the cases of $L>N$ and $L<N$, the abscissa is set from 16 to 128 with step 16.

Note that STE-CV-I and STE-CV-II both compute the estimates iteratively.  As mentioned in \emph{Remark 2}, a rigorous proof of the convergence is unavailable for the {STE-CV-II}. Therefore, we numerically study the convergence of STE-CV-II here. We fix $L=128$ and other parameters are same as those above. 
From Fig. \ref{fig_CG_CF}, we can see the distance $\mathcal{D}^2(\widehat{\mathbf{R}}_{\mathrm{CG},k+1}, \widehat{\mathbf{R}}_{\mathrm{CG},k})$ defined in (\ref{DistanceDefine}) for STE-CV-II decreases with the iterations. Fig. \ref{fig_CG_rho} indicates that,  though STE-CV-II   re-evaluates the cost function at every iteration, the shrinkage factor can still converge to the oracle.

\begin{figure}[!t]
\centering
\includegraphics[width=3.5in]{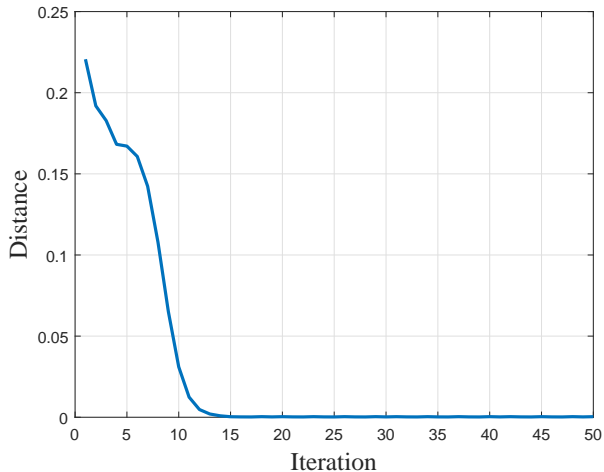}
\caption{The distance $\mathcal{D}^2(\widehat{\mathbf{R}}_{\mathrm{CG},k+1}, \widehat{\mathbf{R}}_{\mathrm{CG},k})$ for STE-CV-II.}
\label{fig_CG_CF}
\end{figure}

\begin{figure}[!t]
\centering
\includegraphics[width=3.5in]{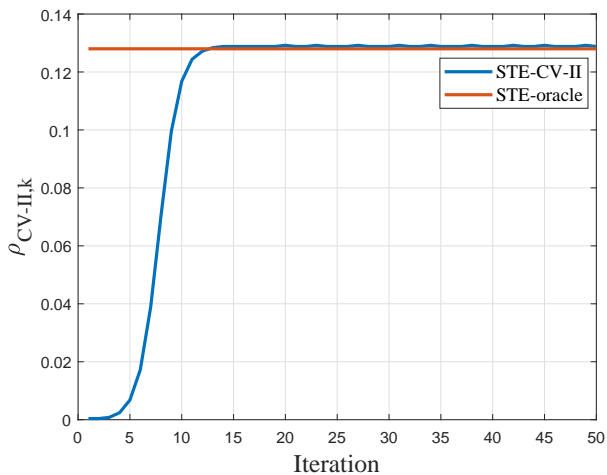}
\caption{The shrinkage factor $\rho_{\mathrm{CV-II},k}$.}
\label{fig_CG_rho}
\end{figure}

\begin{figure}[!t]
\centering
\includegraphics[width=3.5in]{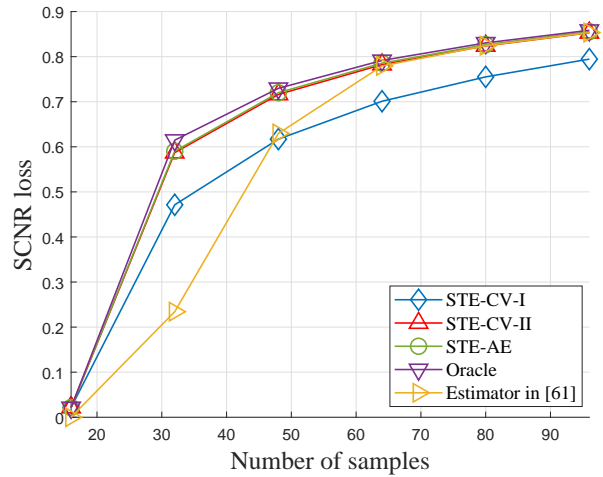}
\caption{SCNR loss versus the number of samples $L$ with compound Gaussian clutter and identity shrinkage target. }
\label{fig_CG_SL1}
\end{figure}

We then compare the proposed estimators {STE-CV-I}, STE-CV-II and STE-AE with the estimator of \cite{8082486}.
These estimators are all iterative, where the initial matrices are set as identity matrices. We also include the STE-oracle estimator defined in (\ref{STEoracle}).
Note that STE-CV-II and STE-AE both re-evaluate the shrinkage factor at each iteration, whereas  STE-CV-I determine  a global shrinkage factor first and then solve the resultant fixed-point equation. 
The STE-oracle estimator measures the upper bound of the SCNR loss for the STE schemes that employ a global shrinkage factor. 
The estimators STE-CV-II, STE-AE  only need an initial matrix for the starting of  iteration, while STE-CV-I firstly assume the true $\mathbf{R}$ is ``known'' and then substitute it using a practical estimator.  
Note that the iteration process for generating the substitution estimator brings additional computational costs. We choose the substituting matrix as the trace-normalized normalized sample covariance matrix (NSCM), i.e.,
$
\widehat{\mathbf{R}}_{\mathrm{NSCM}}=  \frac{N}{L} \sum_{l=1}^{L} \frac{\mathbf{y}_l  \mathbf{y}_l^\mathrm{H}}{\mathbf{y}_l^\mathrm{H}  \mathbf{y}_l},
$ to STE-CV-I to save cost. 
Moreover, STE-CV-I, STE-CV-II, STE-AE each requires only one iterative process for solving the STE fixed-point equation as the choice of the shrinkage factor is either performed beforehand or integrated in the iterative process.

For the application here, the performance metric (\ref{scnrS1}) is also referred to as the SCNR loss. 
Fig. \ref{fig_CG_SL1} shows the SCNR loss versus number of samples $L$. 
For each abscissa, 500 Monte-Carlo experiments are performed. 
It is seen that the proposed methods can achieve the best clutter suppression performance among the estimators considered, especially under limited samples supports.  
The performance of STE-CV-II and STE-AE are very close to that of the oracle estimator and can outperform STE-CV-I. This indicates that iteratively updating the shrinkage factor can improve the performance.

\section{Conclusion}
In this paper, we propose data-driven methods to automatically tune the linear shrinkage factor for $\mathrm{S}^2$CM and STE for Gaussian and compound Gaussian data, respectively.
 The  proposed methods aim to optimize the performance of the MVDR beamformer. 
To achieve this, we adopt the leave-one-out cross-validation (LOOCV) framework with the out-of-sample's output power as the cost and derive results that allow fast calculations of the LOOCV costs and hence low-complexity choice of the shrinkage factors.
For Gaussian samples, we propose an estimator as $\mathrm{S}^2$CM-CV. To obtain a more concise expression, we give a $\mathrm{S}^2$CM-asymptotic estimator ($\mathrm{S}^2$CM-AE).
We also show that, for the case of Gaussian data, the $\mathrm{S}^2$CM-AE is equivalent to a result of \cite{1561576}. 
We then extend  $\mathrm{S}^2$CM-CV and  $\mathrm{S}^2$CM-AE for compound Gaussian data, yielding the STE-CV and STE-AE methods.  
The numerical results demonstrate near-oracle performance of the LOOCV methods for the MVDR beamforming applications considered. 
Future work includes the study of the existence and convergence of STE-CV-II estimator.

\appendices

\section{Proof of Proposition \ref{almostsureequal}}
\label{Pro2}
To lighten the notation, let $g_l(\alpha) \triangleq 1-\frac{N}{L} \phi_l (\alpha)$, $h(\alpha) \triangleq 1-\frac{N}{L} \varphi (\alpha)$, and
\[
\Delta_l^2 (\alpha) \triangleq \frac{ \mathbf{s}^\mathrm{H} \widetilde{\mathbf{R}}_{\mathrm{G}}^{-1}\left( \alpha \right) \mathbf{x}_l\mathbf{x}_l^\mathrm{H}   \widetilde{\mathbf{R}}_{\mathrm{G}}^{-1}\left( \alpha \right) \mathbf{s} }{ \left(\mathbf{s}^\mathrm{H}   \widetilde{\mathbf{R}}_{\mathrm{G}}^{-1}\left( \alpha \right)  \mathbf{s} \right)^2 }.
\]
Then $\mathcal{J}_{\mathrm{CV}} (\alpha)$ and  $\mathcal{J}_{\mathrm{AE}} (\alpha)$ in (\ref{Jcvk1}) and (\ref{Falpha2}) are rewritten as
 \begin{equation}
    \begin{split}
\mathcal{J}_{\mathrm{CV}} (\alpha)=\frac{1}{L} \sum_{l=1}^L \frac{\Delta_l^2 (\alpha)}{g_l^2(\alpha)}, \mathcal{J}_{\mathrm{AE}} (\alpha)=\frac{1}{h^2(\alpha)} \cdot\frac{1}{L} \sum_{l=1}^L \Delta_l^2 (\alpha).
    \end{split}
    \label{CVAEreformulation}
\end{equation}

Recalling  (\ref{CVAEreformulation}), we have 
\begin{equation}
    \begin{split}
  &\left|  \mathcal{J}_{\mathrm{CV}} (\alpha) -\mathcal{J}_{\mathrm{AE}} (\alpha)  \right|\\
  =&\left|    \frac{1}{L} \sum_{l=1}^L (h(\alpha) - g_l(\alpha))  \frac{g_l(\alpha) + h(\alpha)}{g_l^2(\alpha)   h^2(\alpha)} \Delta_l^2 (\alpha)  \right|\\
  \leq &       \left|  \frac{1}{L} \sum_{l=1}^L  \Delta_l^2 (\alpha)  \right|
  \max_{1\leq l\leq L }  \left| h(\alpha) - g_l(\alpha)  \right|  \max_{1\leq l\leq L } \left|  \frac{g_l(\alpha) + h(\alpha)}{g_l^2(\alpha)   h^2(\alpha)} \right|.  
   \end{split}
    \label{diffCVandAE}
 \end{equation}
To proceed with the proof, our strategy is to prove 
 \begin{equation}
    \begin{split}
    \max\limits_{1\leq l\leq L } \left|  \frac{g_l(\alpha) + h(\alpha)}{g_l^2(\alpha)   h^2(\alpha)} \right|< \infty,
\end{split}
    \label{app1cond2}
 \end{equation} 
 \begin{equation}
    \begin{split}
    \left|  \frac{1}{L} \sum_{l=1}^L  \Delta_l^2 (\alpha)  \right| < \infty,
\end{split}
    \label{app1cond1}
 \end{equation} 
 and
\begin{equation}
    \begin{split}
    \sup_{\alpha\in \mathbb{R}^{+}} \max_{1\leq l\leq L }  \left| h(\alpha) - g_l(\alpha)  \right|   \xrightarrow{a.s.} 0.
\end{split}
    \label{diffHandGl}
 \end{equation} 

Therefore, before giving the proof, we give several essential  lemmas under \emph{Assumption} \ref{assuption1}:

\begin{mylem}
\label{Lemma1}
For $\forall \alpha\in \mathbb{R}^{+}$, as $N,L\to \infty$, we have
 \begin{equation}
    \begin{split}
    g_l(\alpha) \in (0,1], h(\alpha)\in (0,1].      
    \end{split}
\end{equation}
\end{mylem}

\emph{Proof}: Since the ranges of $g_l(\alpha)$ and $h(\alpha)$ depend on $\phi_l (\alpha) 
$ and $\varphi(\alpha)$ defined by (\ref{phil}) and (\ref{varphialpha2}), we start with analysing the range of $\phi_l (\alpha)$ and $\varphi(\alpha)$. 
Define $f_{L,l}(\alpha)=\frac{1}{L}\mathbf{x}_l^\mathrm{H} \widetilde{\mathbf{R}}_{\mathrm{G}}^{-1}(\alpha) \mathbf{x}_l$.
 {According to matrix inversion lemma, we have
$
f_{L,l}(\alpha)=\frac{\frac{1}{L}\mathbf{x}_l^\mathrm{H} \widetilde{\mathbf{R}}_{\mathrm{G},l}^{-1}(\alpha)\mathbf{x}_l}{1+\frac{1}{L}\mathbf{x}_l^\mathrm{H} \widetilde{\mathbf{R}}_{\mathrm{G},l}^{-1}(\alpha) \mathbf{x}_l}.
$
Since $\alpha>0$, we have that $\widetilde{\mathbf{R}}_{\mathrm{G},l}(\alpha)$ is positive definite and $\frac{1}{L}\mathbf{x}_l^\mathrm{H} \widetilde{\mathbf{R}}_{\mathrm{G},l}^{-1}(\alpha) \mathbf{x}_l>0$. Therefore, we have $0<f_{L,l}(\alpha)<1$.}
 Since $\widetilde{\mathbf{R}}_{\mathrm{G}}$ is positive definite, we have
\[
 \frac{N}{L} \phi_l (\alpha) \leq f_{L,l}(\alpha) < 1,
0 \le  \frac{N}{L} \varphi (\alpha) =\frac{1}{L} \sum_{l=1}^L f_{L,l}(\alpha)<1.
  \]
We also  have the extra term  in $\phi_l (\alpha)$, i.e.,
\[
\frac{ \left|\mathbf{s}^\mathrm{H}\widetilde{\mathbf{R}}_{\mathrm{G}}^{-1}(\alpha) \mathbf{x}_l \right|^2}{  \mathbf{s}^\mathrm{H}\widetilde{\mathbf{R}}_{\mathrm{G}}^{-1}(\alpha) \mathbf{s}}  \leq 
\mathbf{x}_l^\mathrm{H}\widetilde{\mathbf{R}}_{\mathrm{G}}^{-1}(\alpha) \mathbf{x}_l 
\]
where we utilize the Cauchy-Schwarz inequality, i.e.,
\[\left|\mathbf{s}^\mathrm{H}\widetilde{\mathbf{R}}_{\mathrm{G}}^{-1}(\alpha) \mathbf{x}_l \right|^2 \leq \mathbf{s}^\mathrm{H}\widetilde{\mathbf{R}}_{\mathrm{G}}^{-1}(\alpha) \mathbf{s} \cdot \mathbf{x}_l^\mathrm{H}\widetilde{\mathbf{R}}_{\mathrm{G}}^{-1}(\alpha) \mathbf{x}_l,  \]  on the numerator. Thereby, we have
$\frac{N}{L} \phi_l (\alpha)    \geq 0  $.  $\hfill\blacksquare$

\begin{mycor}
Denote $\lambda_{\mathrm{SCM},\min}$ and $\lambda_{\mathrm{SCM},\max}$ as the minimum and maximum eigenvalue of $\widehat{\mathbf{R}}_{\mathrm{SCM}}$, respectively. Then the quadratic form $\mathbf{s}^\mathrm{H}\widetilde{\mathbf{R}}_{\mathrm{G}}^{-1}(\alpha) \mathbf{s}$ is bounded, i.e.,
\[ \frac{c}{\left(\lambda_{\mathrm{SCM},\max}+\alpha\right)}\leq \frac{1}{L}\mathbf{s}^\mathrm{H}\widetilde{\mathbf{R}}_{\mathrm{G}}^{-1}(\alpha) \mathbf{s} \leq \frac{c}{\left(\lambda_{\mathrm{SCM},\min}+\alpha\right)},\]
a.s., for all large $L$.
\end{mycor}

\emph{Proof}: Recalling $\Vert \mathbf s \Vert^2=N$, this corollary follows easily from the Rayleigh-Ritz theorem, i.e.,
\[ \frac{1}{\lambda_{\mathrm{SCM},\max}+\alpha}\leq \frac{\mathbf{s}^\mathrm{H}\widetilde{\mathbf{R}}_{\mathrm{G}}^{-1}(\alpha) \mathbf{s}}{\mathbf{s}^\mathrm{H}\mathbf{s}} \leq \frac{1}{\lambda_{\mathrm{SCM},\min}+\alpha},\]  
and $L\to \infty$ with $c_n \to c$.
 $\hfill\blacksquare$

\begin{mylem}
\label{Lemma2}
For $\forall \alpha \in \mathbb{R}^{+}$, as $L\to \infty$, $\Delta_l^2 (\alpha)$ can be upperbounded, i.e., $\Delta_l^2 (\alpha)< +\infty$.
\end{mylem}

\emph{Proof}: 
We start with the following fundamental results, which allow for a control of the convergence of the eigenvalue of $\mathbf{R}_{\mathrm{SCM}}$,  under \emph{Assumption} \ref{assuption1}. 
\begin{enumerate}[(1)]
\item The  spectral norm of $\mathbf{R}_{\mathrm{SCM}}$ can be upperbounded, i.e., $\lim \sup_N \Vert \mathbf{R}_{\mathrm{SCM}} \Vert < \infty$, 
which results from \cite[(3.1)]{bai1998} and has been also stated in the proof of \cite[Corollary 2]{COUILLET201499}. This result implies that, there exists $K_c\in ((1+\sqrt{c})^2, \infty)$ such that $\lambda_{\mathrm{SCM},\max} < K_c, $ for all large $L$ a.s.
\item There exists $\epsilon >0$, such that $\lambda_{\mathrm{SCM},\min}>\epsilon, $
for all large $L$ a.s., 
which results from \cite[Lemma 1]{6891244}.
\end{enumerate}
Based on  these results, we then have
 \begin{equation}
    \begin{split}
\Delta_l^2 (\alpha) &\leq \frac{f_{L,l}(\alpha)  }{ \frac{1}{L} \mathbf{s}^\mathrm{H}\widetilde{\mathbf{R}}_{\mathrm{G}}^{-1}(\alpha) \mathbf{s}}<c\left(\lambda_{\mathrm{SCM},\max}+\alpha\right)<\infty,
    \end{split}
    \label{Deltal2}
 \end{equation}
where we utilize the Cauchy-Schwarz inequality on the numerator part of  $\Delta_l^2 (\alpha)$ in the first step in (\ref{Deltal2}). $\hfill\blacksquare$

\begin{mylem}
\label{Lemma3}
For all large $L$, we have
\begin{equation}
    \begin{split}
    \sup_{\alpha\in \mathbb{R}^{+}} \max_{l\leq L }  \left| h(\alpha) - g_l(\alpha)  \right|   \xrightarrow{a.s.} 0.
\end{split}
 \end{equation} 
\end{mylem}

\emph{Proof}: By definition of $g_l (\alpha)$ and $h(\alpha)$, we obtain
\begin{equation}
    \begin{split}\nonumber
    &  \left| h(\alpha) - g_l(\alpha)  \right| = \left|  \left(  f_{L,l}(\alpha)- p_l (\alpha)  \right)- \frac{1}{L} \sum_{l=1}^L f_{L,l}(\alpha)       \right|,
    \end{split}
 \end{equation}
where
$
p_l (\alpha)=\frac{1}{L}\frac{ \left|\mathbf{s}^\mathrm{H}\widetilde{\mathbf{R}}_{\mathrm{G}}^{-1}(\alpha) \mathbf{y}_l \right|^2}{  \mathbf{s}^\mathrm{H}\widetilde{\mathbf{R}}_{\mathrm{G}}^{-1}(\alpha) \mathbf{s}}\geq 0. 
$

Define \[\bar{f}(\alpha) \triangleq  \frac{c_N e_N (-\alpha)}{1+c_N e_N (-\alpha)} \]
with $e_N (z)$ the unique positive solution of 
\[
e_N (z)=\int  \frac{1}{(1+c_N e_N (z))^{-1}   t  -z}dF^{\mathbf{R}}(t),
\]
where $F^{\mathbf{R}}(t)$ is the eigenvalue distribution of $\mathbf{R}$ \cite{6891244,bai1998}. 
Then we have
\begin{equation}
    \begin{split}\nonumber
&
\max_{l\leq L}\left| h(\alpha) - g_l(\alpha)  \right| \\
 \leq  & \frac{1}{L} \sum_{l=1}^L  \left| f_{L,l}(\alpha) -\bar{f}(\alpha)   \right|           + \max_{l\leq L}  \left|  f_{L,l}(\alpha)-\bar{f}(\alpha) \right|       +   \max_{l\leq L}   p_l (\alpha)  \\
 \leq  & 2\left(\max_{l\leq L}  \left|  f_{L,l}(\alpha)-\bar{f}(\alpha) \right|    \right)   +   \max_{l\leq L}   p_l (\alpha) .
\end{split}
\end{equation}
Note first that \cite[(10)]{6891244} shows that \[\max_{l\leq L} \left|  f_{L,l}(\alpha)-\bar{f}(\alpha) \right| \xrightarrow{a.s.} 0.\]
Then, by definition of mathematical expectation, we have, for all $\alpha \in \mathbb{R}^{+}$, 
\begin{equation}
    \begin{split}\nonumber
    \mathbb{E} (p_l (\alpha))&=\frac{1}{L}\sum_{l=1}^L  p_l (\alpha) =\frac{1}{L} \frac{ \mathbf{s}^\mathrm{H} \widetilde{\mathbf{R}}_{\mathrm{G}}^{-1}\left( \alpha \right) \mathbf{R}_{\mathrm{SCM}}  \widetilde{\mathbf{R}}_{\mathrm{G}}^{-1}\left( \alpha \right) \mathbf{s} }{ \mathbf{s}^\mathrm{H}   \widetilde{\mathbf{R}}_{\mathrm{G}}^{-1}\left( \alpha \right)  \mathbf{s}  }\\
    &\leq  \frac{1}{L} \cdot \frac{\lambda_{\mathrm{SCM},\max}}{\lambda_{\mathrm{SCM},\max}+\alpha}<\frac{1}{L} ,
    \end{split}
\end{equation}
a.s., for all large $L$.
Then we have for all $\zeta>0$, 
\begin{equation}
    \begin{split}\nonumber
    \sum_{l=1}^L P( p_l (\alpha)  >\zeta  ) \leq \sum_{l=1}^L \frac{ \mathbb{E} (p_l (\alpha))}{\zeta} <\frac{1}{\zeta}<\infty,
    \end{split}
\end{equation}
where we utilize the Markov inequality in  the first step. 
The Borel Cantelli lemma therefore ensures, for all $\alpha \in \mathbb{R}^{+}$, 
$\max_{l\leq L}  p_l (\alpha)  \xrightarrow{a.s.} 0. $
We then conclude that
\begin{equation}
    \begin{split}\nonumber
    &\sup_{\alpha\in \mathbb{R}^{+}} \max_{l\leq L }  \left| h(\alpha) - g_l(\alpha)  \right| \\
    &\leq  \sup_{\alpha\in \mathbb{R}^{+}} 2 \left(\max_{l\leq L }  \left|  f_{L,l}(\alpha)-\bar{f}(\alpha) \right|   \right)    +   \sup_{\alpha\in \mathbb{R}^{+}} \max_{l\leq L }   p_l (\alpha) ,\\
    &\xrightarrow{a.s.} 0,
\end{split}
 \end{equation}
which completes the proof. $\hfill\blacksquare$
 
Then (\ref{app1cond2}), (\ref{app1cond1}) and (\ref{diffHandGl}) are obtained by \emph{Lemma 1}, \emph{2} and \emph{3}, respectively.  Then the proof of \emph{Proposition 1} is completed.

\ifCLASSOPTIONcaptionsoff
  \newpage
\fi

\end{document}